\documentclass[12pt,a4paper]{article}
\pdfoutput=1
\usepackage{graphicx}
\usepackage{tikz}
\usetikzlibrary{decorations.pathreplacing,decorations.markings,snakes}
\usetikzlibrary{decorations.pathmorphing}
\tikzset{snake it/.style={decorate, decoration=snake}}
\usepackage{jheppub}
\usepackage{amsmath,amssymb,euscript,array,mathrsfs,appendix,ctable}
\usepackage{bm}
\usepackage[small]{caption}
\usepackage{xcolor}
\usepackage{float}
\usepackage{braket}
\usepackage{comment}
\usepackage{xcolor}
\usepackage{empheq}
\definecolor{lightblue}{RGB}{100,180,255}
\usepackage{mathtools}
\newcommand \arXiv [1]{\href{http://arxiv.org/abs/#1}{\tt arXiv:#1}}

\def\ben
{\begin{equation}}
\def\een{\end{equation}}

    \let\r=\rho

    \let\L=\Lambda
   
 \let\W=\mu

\def\W={\cal W}
\def\L ={\cal L}

\def\be{\begin{equation}}
\def\ee{\end{equation}}
\def\ba{\begin{array}}
\def\ea{\end{array}}

\def\dalemb#1#2{{\vbox{\hrule height .#2pt
        \hbox{\vrule width.#2pt height#1pt \kern#1pt
                \vrule width.#2pt}
        \hrule height.#2pt}}}

\newcommand{\bea}{\begin{eqnarray}}
\newcommand{\eea}{\end{eqnarray}}
\newcommand{\tr}{{\rm tr} }

\title{Evaporating Black Hole Interior and Complexity Evolution}
\author{Nicol\`o Bragagnolo and S. Prem Kumar}
\affiliation{Centre for Quantum Fields and Gravity, \\Department of Physics,\\
Swansea University,\\
Singleton Park, Swansea,\\
SA2 8PP, U.K.}
\emailAdd{nicolo.bragagnolo@swansea.ac.uk, s.p.kumar@swansea.ac.uk}
\abstract{
We study the evolution of the interior of an evaporating black hole in a simple model of Jackiw-Teitelboim (JT) gravity with an end-of-the-world (EoW) brane, where evaporation is modeled by entangling the brane's internal states with an auxiliary radiation system. To probe the black hole interior, we consider a geodesic length extracted from a boundary-to-brane two-point function and interpret its renormalised value as a measure of subsystem complexity. Our computation, based on quenched disorder averaging,  includes non-perturbative gravitational effects from both spacetime wormholes and replica wormholes, encoding ensemble averaging over the dual random Hamiltonian and  brane-state couplings. Unlike non-evaporating black holes, for which complexity first grows linearly and then plateaus at late times $\sim{\cal O}(e^{S_{\rm BH}})$, we find that complexity evolution of the black hole subsystem in the  evaporating case differs drastically, depending nontrivially on the  dimension of the emitted radiation Hilbert space. It grows linearly at early times, reaches a maximum shortly before the Page-time crossover $\sim{\cal O}({S_{\rm BH}})$, and then decays exponentially. We further show that the relative fluctuations of the interior length remain small before the Page time but become of order one and eventually large at later times: this signals a loss of self-averaging, with the ensemble-averaged complexity dominated by rare configurations rather than by typical realisations.
}

\begin{document}
\maketitle
\flushbottom
\section{Introduction}
The entanglement structure of an evaporating black hole with its emitted Hawking radiation is expected to undergo nontrivial evolution past the Page time of the black hole \cite{Hawking:1975vcx, Hawking:1976ra, Page:1993wv, Page:1993df}. Within semiclassical gravity, the corresponding changes in entanglement measures are dictated by the appearance of entanglement islands behind the horizon, demarcated by quantum extremal surfaces, accounting for the fact that the semiclassical modes behind the horizon have been radiated away and are in fact contained in the radiation Hilbert space \cite{Penington:2019npb, Penington:2019kki, Almheiri:2019qdq}.
The semiclassical replica wormhole saddle points underlying this  picture tell us that insofar as measures of entanglement such as von Neumann and  R\'enyi entropies are concerned, the magnitude of the fluctuations around the classical black hole background, or the Hawking saddle, becomes large and dominates the contributions to the respective entropies of entanglement. 

It is not, however, immediately clear from these calculations, whether there are significant departures from the  classical spacetime geometry behind the horizon in observables unrelated to measures of entanglement. 

In this paper we attempt to address this point by examining the volume of the interior of an evaporating black hole within a simple toy model of evaporation which was originally introduced in \cite{Penington:2019kki}. Our black hole is described within the effective two-dimensional Jackiw-Teitelboim (JT) gravity \cite{Teitelboim:1983ux, Jackiw:1984je, Maldacena:2016upp} which captures near extremal black hole dynamics. In order to model the entanglement of the black hole with  its Hawking quanta, the latter are introduced as an auxiliary reference system $R$ whose states are  maximally entangled with partner states of an end of the world (EoW) brane behind the horizon (see Figure \ref{fig1}),
\be
    \ket{\Psi}=\frac{1}{\sqrt{k}}\sum_{i=1}^{k}\ket{\psi_{i}}_{B}\ket{i}_{R}\,,
\ee
where $|\psi_i\rangle_B$ is the state of the black hole $B$ with EoW brane in the state $i$, and $|i\rangle_R$ is the state of the auxiliary ``radiation" system. As the evaporation proceeds, $k$, the dimension of the radiation Hilbert space, grows and the na\"ive or coarse-grained thermal entropy $S_{\rm rad}$ tracks this growth as  $k(t)= e^{S_{\rm rad}(t)}$.

Within this 2d model we will focus attention on a measure of the volume  of the black hole interior characterised by the expectation value of a length operator defined via the correlation function of a probe bulk field $O_\Delta$ with one insertion at the EoW brane and another at the AdS$_2$ boundary \cite{Alishahiha:2022kzc},
\be
\label{proposal-geodesic-unequipped}
\braket{L(t)}=-\lim_{\Delta\to0}\frac{\partial\braket{O_\Delta\left(\frac{\beta}{2}+it\right)\,O_\Delta(0)}}{\partial\Delta}\,. 
\ee
This definition closely parallels that of the length of the Einstein-Rosen bridge in the two-sided AdS$_2$ black hole  \cite{Iliesiu:2021ari}. Here $\Delta$ denotes the conformal dimension of the boundary operator dual to the bulk operator $O_\Delta$. The definition \eqref{proposal-geodesic-unequipped} follows intuitively from the observation that the bulk-to-boundary propagator in AdS$_2$ is $\sim e^{-\Delta L}$ where $L$ is the geodesic distance between the bulk and boundary points.
Other prescriptions could in principle be adopted to define the interior length, however  \eqref{proposal-geodesic-unequipped} is a natural one, and has been used  to understand the expected behaviour of the quantum complexity of the black hole state or equivalently that of its holographic dual system \cite{Susskind:2014rva, Susskind:2015toa, Brown:2016wib, Susskind:2020wwe, Balasubramanian:2019wgd}. In the present context, the interior length of the evaporating black hole, appropriately renormalised, must be interpreted as a \textit{subsystem} complexity \cite{Alishahiha:2015rta, Chen:2018mcc, Agon:2018zso, Carmi:2016wjl}.

A distinctive feature of the model we study is that it admits a dual random-matrix description
involving both a random Hamiltonian and random couplings specifying the brane states \cite{Penington:2019kki}. Correspondingly, the relevant two-point function is obtained from the $n$-replicated theory, to implement  a {\em quenched disorder average}  of the free energy \cite{Aharony:2018mjm, Engelhardt:2020qpv} in the presence of sources for local operators (cf. eq. \eqref{replica_result})
and the corresponding geodesic length is extracted by taking the $\Delta\to 0$ derivative as dictated by eq.\,\eqref{proposal-geodesic-unequipped}.
As a result, the computation of the interior volume receives two conceptually
different non-perturbative contributions: ordinary spacetime wormholes, associated with the genus expansion of
the pure JT gravitational path integral, and replica wormholes, which arise from
the ensemble average over the brane data and encode the correlations responsible
for the Page transition. 

\begin{figure}[ht]
    \centering
    \includegraphics[width=0.7\linewidth]{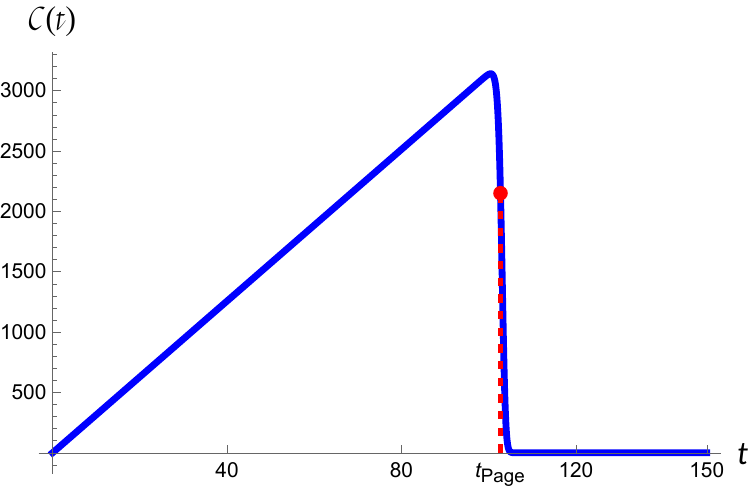}
    \caption{The renormalised geodesic length ${\cal C}(t)=\overline{\braket{L_{k(t)}(t)}}-\overline{\braket{L_{k(t)}(0)}}$ which we identify with the subsystem complexity of the black hole state  in the canonical ensemble. We have set $\beta=0.2$, $S_{0}=8$ and $\dot{S}_{\text{rad}}=2$ in eq.\,\eqref{comp_canonical} to obtain this plot. The crossover (inflection point) occurs at Page time $t_{\text{Page}}\simeq 102.7$. Note that we have kept $\beta$ fixed, while in a realistic setting $\beta$ must evolve adiabatically with $t$ as well.}
    \label{length_canonical}
\end{figure}
Our main result is that the interior length as defined above, after appropriate regularisation, interpreted as  complexity of an evaporating black hole, behaves
qualitatively differently from that of the non-evaporating AdS black hole. The evaluation of the length is controlled in part by the non-perturbative spectral two-point function of JT gravity \cite{Saad:2019lba, Alishahiha:2022kzc, Iliesiu:2021ari, Saad:2019pqd, Blommaert:2020seb}.

Crucially, we find a new nontrivial ingredient in the evaporating case from summing over all replica wormhole contributions, which enters as a kernel in the energy integral that determines the non-perturbative length. The new resummed contributions, denoted ${\cal I}(n)$ in the $n$-replicated theory, arise from summing over all weighted partitions of replica geometries into connected and disconnected multi-boundary components. ${\cal I}(n)$ turns out to be a non-monotonic function of $e^{S_{\rm BH}}/k$, the ratio of the dimensions of the semiclassical black hole and radiation Hilbert  spaces.
In the microcanonical ensemble, the relevant kernel  admits a probabilistic representation,
\begin{equation}
\left.\partial_n {\cal I}(n)\right|_{n=0}
=
\mathbb{E}_{X}
\left[\frac{X}{(X+1)^2}\right]\,, \qquad\qquad \text{with}\;\; X\sim{\rm Poisson}\left(e^{S_{\text{BH}}(t)-S_{\text{rad}}(t)}\right)\,.
\end{equation}
It can be interpreted as the expectation value of the function $f(X)=X/(X+1)^2$ of a Poisson distributed random variable $X$ with mean value $e^{S_{\rm BH}}/k$. 
This function has a maximum when $S_{\rm BH}= S_{\rm rad} +{\cal O}(1)$,  making the Page-time crossover especially transparent. 
We find, both in canonical and microcanonical descriptions,  that
the renormalised interior length grows linearly at early times and reaches a maximum at a time
parametrically close to the Page time. Explicitly, the time evolution of the renormalised interior length can be summarised as,
\begin{equation}
\overline{\braket{L_{k(t)}(t)}}_{\rm ren}\,\sim
\begin{cases}
t \,, & t\ll t_{\rm Page}\,,\\[4pt]
e^{2\left(S_{\text{BH}}(t)-S_{\rm rad}(t)\right)}\, t \,, & t_{\rm Page}\ll t\lesssim e^{S_0}\,,
\end{cases}
\label{eq:intro_micro_scaling}
\end{equation}
with a further late-time suppression beyond the non-perturbative scale $t\sim e^{S_0}$. Here $S_0$ is the extremal or zero temperature entropy.
Thus, unlike the non-evaporating case where the complexity grows linearly until plateauing at $t\sim e^{S_0}$, the
ensemble-averaged complexity of the evaporating black hole turns over near
Page time and then decreases exponentially. In the high temperature limit, our result for complexity evolution in the canonical ensemble is particularly simple,
\be
{\mathcal{C}}(t)
    =\,\frac{2\pi}{\beta}\,t\,\,\frac{e^{2S_{\rm BH}(\beta)}}{\left(k(t)\,+\,e^{S_{\rm BH}(\beta)}\right)^{2}}\,,
\ee
and is depicted in Figure \ref{length_canonical} for $k(t)=e^{\dot{S}_{\rm rad} t}$.

A second important result concerns fluctuations. The variance of the interior length can be extracted from the expectation value of the squared geodesic length, and we find  the relative magnitude of the fluctuation to be,
\begin{equation}
\frac{\sigma_{\bar L}(t)}{\overline{\braket{L_{k(t)}(t)}_{\rm ren}}}
\sim
\left(\overline{\braket{L_{k(t)}(t)}}_{\rm ren}\right)^{-1/2} \, .
\end{equation}
This remains parametrically small before Page time,
so the ensemble average faithfully captures the behaviour of a typical realization. After the peak, however, the mean length decays exponentially and the relative
fluctuations cease to be small: they become order one shortly after the Page-transition time and then grow exponentially. In this sense, the post-Page time regime is characterised
by a loss of self-averaging.

The rest of this paper is organised as follows. In section \ref{sec2} we review the JT gravity setup with
an EoW brane, and the evaporation model obtained by entangling the brane states with
an auxiliary radiation system. In section \ref{sec3} we explain the replica prescription for the
ensemble-averaged boundary-to-brane correlator and its interpretation as a complexity
observable. Section \ref{sec4} reviews the canonical quantization of JT gravity with an EoW
brane. Section \ref{sec5} contains the replica computation of the ensemble-averaged geodesic
length. In sections \ref{microcanonical} and \ref{canonical} we study the complexity in the microcanonical and canonical
ensembles, respectively, and derive the associated variance. We conclude in section \ref{summary}
with a discussion of the results and future directions. An extensive appendix is devoted to the technical steps involved in a number of analytical derivations.

\section{Setup: JT gravity with EoW brane}
\label{sec2}
Let us review the simple $2$d gravity model for an evaporating black hole originally introduced in \cite{Penington:2019kki} to derive the Page curve by summing over replica geometries with different topologies. The idea is to consider a two-sided black hole in JT gravity \cite{Teitelboim:1983ux, Jackiw:1984je, Maldacena:2016upp}, with the geometry cut off by an  “end-of-the-world brane" (EoW brane) behind the horizon with brane tension $\mu$, which can be viewed as a particle of mass  $\mu$ \cite{Kourkoulou:2017zaj}.
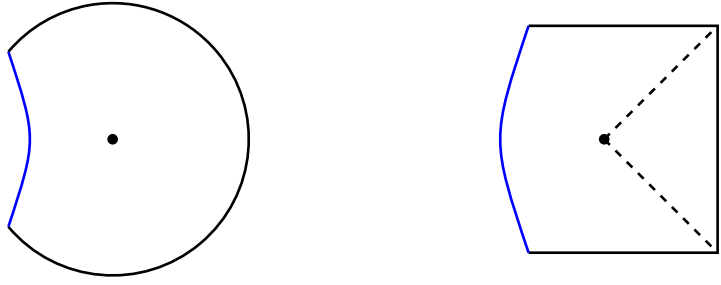
\begin{figure}[h]
\centering
\begin{tikzpicture}
    \def\r{1.8}

    \coordinate (A) at (-140:\r);
    \coordinate (B) at (140:\r);

    \draw[line width=1pt] (A) arc[start angle=-140, end angle=140, radius=\r];

    \fill (0,0) circle (2pt);

    \draw[blue, line width=1pt]
        (A) .. controls (-1,0) .. (B);

    \coordinate (S) at (5.5, -1.5); 

    \def\L{3} 

    \draw[line width=1pt]
        (S) -- ++(2.5,0) -- ++(0,\L) -- ++(-2.5,0); 

    \draw[blue, line width=1pt]
        (S) .. controls (5,0) .. (5.5,1.5);

    \fill (6.5,0) circle (2pt);

    \draw[black, dashed, line width=1pt]
        (6.5,0)--(8,-1.5);
    \draw[black, dashed, line width=1pt]
        (6.5,0)--(8, 1.5);
\end{tikzpicture}
\caption{Euclidean (left) and Lorentzian (right) geometries of a black hole with an EoW brane (depicted in blue) behind the horizon.}
\label{fig1}
\end{figure}

The overall Euclidean action of the model is given by
\begin{equation}\label{eq:overallaction}
    I_{E}= I_{\text{JT}} + I_{\rm EoW}\,
\end{equation}
where the pure JT gravity action including boundary terms is
\begin{equation}\label{eq:EuclideanJTaction}
    I_{\text{JT}}=-\frac{S_0}{4\pi}\left( \int_{\mathcal{M}} \sqrt{g}\,R + 2 \int_{\partial \mathcal{M}} \sqrt{h} \,K \right) - \int_{\mathcal{M}} \sqrt{g}\, \phi \left(R+2\right)
- 2 \int_{\partial \mathcal{M}} \sqrt{h} \,\phi \left(K-1\right)\,,
\end{equation}
with $S_0/4\pi$ the extremal entropy of the AdS$_2$ black hole.
The action of the EoW brane with tension $\mu$ is of the form
\begin{equation}\label{eq:EoWbraneaction}
    I_{\rm EoW}=\mu \int_{\rm EoW}ds\,. 
\end{equation}
As is well known, the equation of motion for the dilaton forces the geometry to be locally AdS$_2$, whilst varying with respect to metric yields a nontrivial equation for the dilaton,
\begin{equation}
    R+2=0\,,\qquad \nabla_{\mu}\nabla_{\nu}\phi - g_{\mu \nu} \nabla^2 \phi + g_{\mu \nu} \phi=0\,.
\end{equation}
In addition, both the induced metric and the dilaton value at the asymptotic AdS boundary are fixed as,
\begin{equation}
  ds^2|_{\partial \mathcal{M}} =\frac{d\tau^2}{\epsilon^2}\,,\;\;\;\;\;\;\;\;\;\;\;\;
    \phi|_{\partial \mathcal{M}} = \frac{\phi_b}{\epsilon}\,.
\end{equation}
Here $\tau \sim \tau + \beta$ can be interpreted as the periodic Euclidean time of the holographic boundary dual to the JT gravity system and $\epsilon$ parametrises the (UV) cutoff scale, the AdS$_2$ boundary approached in the  $\epsilon \rightarrow 0$ limit. For simplicity, we will set the constant $\phi_{b}=1$.

Two further boundary conditions must be imposed at  the EoW brane,
\begin{equation}
    K\big|_{\rm EoW} = 0\,,\;\;\;\;\;\;\;\;\;\; \partial_{n}\phi\big|_{\rm EoW} = \mu\,,
\end{equation}
where $\partial_n$ denotes the derivative normal to the EoW brane and $K$ the extrinsic curvature. 
The vanishing of the extrinsic curvature implies that the EoW brane is a geodesic, whereas the second condition $\partial_{n}\phi = \mu$ results in vanishing brane Hamiltonian $H_{\text{brane}}$, consistently with gravity being dynamical on the EoW brane \cite{Gao:2021uro}.

\subsection{Equipping the EoW  brane}
Up to now, the EoW brane is just the geodesic of a particle of mass $\mu$ propagating behind the horizon, yielding a $\mathbb{Z}_{2}$ quotient of an ordinary two-sided black hole \cite{Kourkoulou:2017zaj}.

To model evaporation, we endow the EoW brane with an internal Hilbert space of dimension $k$, spanned by a set of orthogonal states labeled by $i=1,\ldots,k$ \cite{Penington:2019kki}. These degrees of freedom represent modes in the black hole interior correlated with the Hawking radiation. Evaporation is then mimicked by entangling the brane degrees of freedom with states of an auxiliary radiation system $R$. The combined state takes the form
\begin{equation}
    \ket{\Psi}=\frac{1}{\sqrt{k}}\sum_{i=1}^{k}\ket{\psi_{i}}_{B}\ket{i}_{R}\,,
\end{equation}
where $\ket{\psi_{i}}_{B}$ is the state of the black hole $B$ with the EoW brane in the internal state with flavour $i$, and $\ket{i}_{R}$ is the entangled state of the auxiliary radiation system $R$.

This model, originally introduced in \cite{Penington:2019kki}, captures the essential ingredients needed to recover the Page curve of the entropy of the radiation system $R$ using the replica trick. The most direct way to show it is via the study of the purity of the radiation system $R$. For this, we first introduce the reduced density matrix $\rho_{R}$ of the radiation system,
\begin{equation}
    \rho_{R}=\frac{1}{k}\sum_{i,j=1}^{k}\ket{j}\bra{i}_{R}\langle\psi_{i}|\psi_{j}\rangle_{B}
\end{equation}
whose matrix elements $\langle\psi_{i}|\psi_{j}\rangle$ are amplitudes between semiclassical gravity states, subject to  boundary conditions depicted in Figure \ref{bc-gampl}.
\begin{figure}[h]
\centering
\begin{tikzpicture}
    \draw[black, line width=1pt, postaction={decorate},
        decoration={
            markings,
            mark=at position 0.5 with {
                \arrow[scale=1.5]{stealth reversed}
            }
        }]
        (0,0) .. controls (1,-0.5) and (2,0.5) .. (3,0);
    \draw[blue, dashed, line width=1pt]
        (0,0)--(-1,0) node[left] {$i$};
    \draw[blue, dashed, line width=1pt]
        (3,0)--(4,0) node[right] {$j$};
\end{tikzpicture}
\caption{\footnotesize Boundary data entering the amplitude $\langle\psi_{i}|\psi_{j}\rangle$. The solid curve denotes the asymptotic boundary of renormalised length $\beta$, with the arrow fixing the ket-to-bra orientation. Dashed segments indicate EoW-brane endpoints carrying the flavour labels $i$ and $j$.}
\label{bc-gampl}
\end{figure}

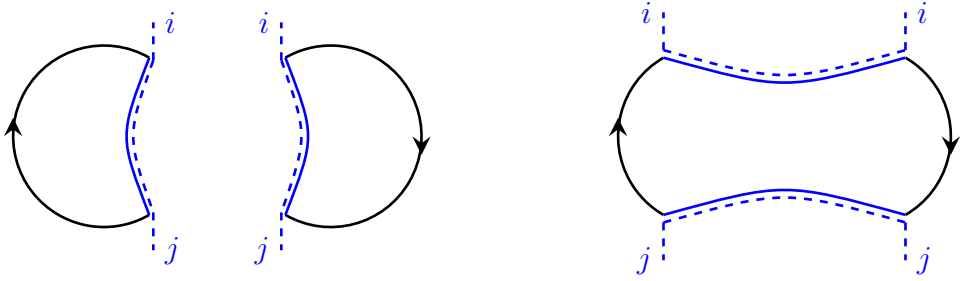
\begin{figure}[H]
\centering
\begin{tikzpicture}
     \draw[black, line width=1pt, postaction={decorate},
        decoration={
            markings,
            mark=at position 0.5 with {
                \arrow[scale=1.5]{stealth reversed}
            }
        }] (0,0) ++(60:1.2cm) arc [start angle=60, end angle=300, radius=1.2cm];
        \draw[blue, line width=1pt] ({1.2*cos(60)}, {1.2*sin(60)} ).. controls (0.2,0)..( {1.2*cos(60)}, { -1.2*sin(60)});
        \draw[blue, dashed, line width=1pt] ({1.2*cos(57)}, {1.2*sin(57)} ).. controls (0.3,0)..( {1.2*cos(57)}, { -1.2*sin(57)});
        \draw[blue, dashed, line width=1pt] ({1.2*cos(57)}, {1.2*sin(57)} )--( {1.2*cos(57)}, {1.2*sin(57)+0.5}) node[right] {$i$};
        \draw[blue, dashed, line width=1pt] ({1.2*cos(57)}, {-1.2*sin(57)} )--( {1.2*cos(57)}, {-1.2*sin(57)-0.5}) node[right] {$j$};

    \draw[black, line width=1pt, postaction={decorate},
        decoration={
            markings,
            mark=at position 0.5 with {
                \arrow[scale=1.5]{stealth reversed}
            }
        }] (3,0) ++(-120:1.2cm) arc [start angle=-120, end angle=120, radius=1.2cm];
        \draw[blue, line width=1pt] ({3+1.2*cos(120)}, {1.2*sin(120)} ).. controls (2.8,0)..({3+1.2*cos(120)}, {-1.2*sin(120)});
        \draw[blue, dashed, line width=1pt] ({3+1.2*cos(123)}, {1.2*sin(123)} ).. controls (2.7,0)..({3+1.2*cos(123)}, {-1.2*sin(123)});
        \draw[blue, dashed, line width=1pt] ({3+1.2*cos(123)}, {1.2*sin(123)} )--( {3+1.2*cos(123)}, {1.2*sin(123)+0.5}) node[left] {$i$};
        \draw[blue, dashed, line width=1pt] ({3+1.2*cos(123)}, {-1.2*sin(123)} )--( {3+1.2*cos(123)}, {-1.2*sin(123)-0.5}) node[left] {$j$};


    \draw[black, line width=1pt, postaction={decorate},
        decoration={
            markings,
            mark=at position 0.5 with {
                \arrow[scale=1.5]{stealth reversed}
            }
        }] (8,0) ++(120:1.2cm) arc [start angle=120, end angle=240, radius=1.2cm];

    \draw[black, line width=1pt, postaction={decorate},
        decoration={
            markings,
            mark=at position 0.5 with {
                \arrow[scale=1.5]{stealth reversed}
            }
        }] (10,0) ++(-60:1.2cm) arc [start angle=-60, end angle=60, radius=1.2cm];

    \draw[blue, line width=1pt] ({8+1.2*cos(120)}, {1.2*sin(120)} ).. controls (9,0.6)..({10+1.2*cos(60)}, {1.2*sin(60)});

    \draw[blue, dashed, line width=1pt] ({8+1.2*cos(120)}, {0.1+1.2*sin(120)} ).. controls (9,0.7)..({10+1.2*cos(60)}, {0.1+1.2*sin(60)});

    \draw[blue, dashed, line width=1pt] ({8+1.2*cos(120)}, {0.1+1.2*sin(120)} )--({8+1.2*cos(120)}, {0.6+1.2*sin(120)} ) node[left] {$i$};

    \draw[blue, dashed, line width=1pt] ({10+1.2*cos(60)}, {0.1+1.2*sin(60)} )--({10+1.2*cos(60)}, {0.6+1.2*sin(60)} ) node[right] {$i$};

    \draw[blue, line width=1pt] ({8+1.2*cos(120)}, {-1.2*sin(120)} ).. controls (9,-0.6)..({10+1.2*cos(60)}, {-1.2*sin(60)});
    
    \draw[blue, dashed, line width=1pt] ({8+1.2*cos(120)}, {-0.1-1.2*sin(120)} ).. controls (9,-0.7)..({10+1.2*cos(60)}, {-0.1-1.2*sin(60)});

    \draw[blue, dashed, line width=1pt] ({8+1.2*cos(120)}, {-0.1-1.2*sin(120)} )--({8+1.2*cos(120)}, {-0.6-1.2*sin(120)} ) node[left] {$j$};

    \draw[blue, dashed, line width=1pt] ({10+1.2*cos(60)}, {-0.1-1.2*sin(60)} )--({10+1.2*cos(60)}, {-0.6-1.2*sin(60)} ) node[right] {$j$};

\end{tikzpicture}
\caption{\footnotesize Depiction of the two possible ways of filling in the bulk the boundary conditions of $|\langle\psi_{i}|\psi_{j}\rangle|^{2}$. In order to compute the purity, we sum over the indices $i,j$ by connecting the dashed lines jointly.}
\label{puritycontr}
\end{figure}

The purity of the radiation system $R$ is, by definition,
\begin{equation}
\tr(\rho_{R}^2) = \frac{1}{k^2}\sum_{i,j = 1}^k|\langle \psi_i|\psi_j\rangle|^2.\label{purityFormula}
\end{equation}
The gravitational evaluation of \eqref{purityFormula} requires summing over bulk geometries compatible with the boundary conditions of the two copies of the amplitude. At leading order there are two topologically distinct contributions, shown in Figure \ref{puritycontr}. One consists of two disconnected discs, while the other is a single connected disc joining the replicas and therefore gives the replica-wormhole contribution.\\
We are going to use the notation $Z_n = Z_n(\beta)$ to represent the gravitational path integral on a disc-topology geometry with a boundary consisting of alternating segments of $n$ physical boundaries of renormalised lengths $\beta$, and $n$ EoW branes. Using this notation, we can evaluate the sum of the two contributions of the purity as
\begin{equation}
\tr(\rho_{R}^2) =  \frac{kZ_1^2 + k^2Z_2}{(k Z_1)^2} = \frac{1}{k} + \frac{Z_2}{Z_1^2}.\label{visible}
\end{equation}
The two terms in the numerator of \eqref{visible} have a direct topological interpretation. For the disconnected saddle, contraction of the EoW-brane indices produces a single index loop, giving a factor $k$, while the gravitational path integral factorises into two copies of $Z_1$. For the connected saddle there are two independent brane-index sums, giving $k^{2}$, together with a single connected gravitational contribution $Z_2$. The denominator supplies the normalisation by ${\rm tr}(\rho_R)^2$.

The leading qualitative behaviour of the purity can be captured by retaining only the topological $S_0$ term in the JT gravity action. This term weights the contribution of each topology with Euler characteristic $\chi$,  by $e^{S_0\chi}$. Since the topology involved in $Z_n$ is disc-like for any value of $n$, and $\chi = 1$ for the disc, we will have
\begin{equation}
Z_n \,\propto \,e^{S_0}\,.
\end{equation}
Using this formula, \eqref{visible} can be written schematically as
\begin{equation}
\tr(\rho_{R}^2) = k^{-1} + c\,e^{-S_0}\;,\label{purity}
\end{equation}
for some energy and temperature dependent constant $c$. Equation \eqref{purity} makes the Page transition transparent. When $k\ll e^{S_0}$, the $k^{-1}$ term is dominant and the purity is controlled by the disconnected saddle. As $k$ increases, this contribution decreases until it becomes comparable to the connected contribution of order $e^{-S_0}$. For $k\gtrsim e^{S_0}$, the connected replica-wormhole saddle therefore controls the purity, which becomes independent of $k$ at leading order. The change in the dominant topology is what causes the radiation Rényi entropy to saturate at the black-hole entropy scale rather than continuing to grow with the radiation Hilbert-space dimension.

Using combinatorial arguments it is possible to get the precise evolution of any $n$-Rényi entropy, including the radiation entanglement entropy. The details of the treatment in the  microcanonical ensemble can be found in Appendix \ref{SR}.

\section{Black hole volume/complexity prescription}
\label{sec3}
It turns out that pure JT gravity with EoW branes equipped in the way described above is dual to a boundary ensemble of theories, defined by a random Hamiltonian $H$ and a set of brane states $\ket{\psi_{i}(\beta)}$, which satisfy \cite{Penington:2019kki}
\begin{equation}\label{psi-random}
    \ket{\psi_{i}(\beta)}=\sum_{a}\;\Gamma\left(\mu+\frac{1}{2}+i\sqrt{2E_{a}}\right)e^{-\frac{\beta E_{a}}{2}}C_{i,a}\ket{E_{a}}\,,
\end{equation}
where $\ket{E_{a}}$ are the eigenstates of the Hamiltonian $H$ and the coefficients $C_{i,a}$ are independent and identically distributed (i.i.d.) complex Gaussian random variables. In the dual random matrix ensemble of this model, the Hamiltonian $H$ can be viewed as a $L\times L$ matrix and $C$ as a $k\times L$ matrix, since the energy and flavor indices assume values $a\in\{1,\dots,L\}$ and $i\in\{1,\dots,k\}$ respectively.  The factor of $\Gamma(\mu+\frac12 + i\sqrt{2 E_a})$ originates from an integral over the EoW brane wave function and the Hartle-Hawking state in the geodesic  basis \cite{Penington:2019kki, Yang:2018gdb} which we will review below. 

\subsection{Length from correlators}
 A prescription for extracting the interior geodesic length in pure JT gravity was introduced in \cite{Iliesiu:2021ari}. The construction sums over non-self-intersecting geodesics connecting operator insertions and remains applicable when higher-topology contributions to the gravitational path integral are included. Introducing $\Delta$ as a regulator, the length is defined by, 
\begin{equation}
    \braket{L}=\lim_{\Delta\to 0}\left\langle\sum_{\gamma}L_{\gamma}e^{-\Delta L_{\gamma}}\right\rangle\,,
\end{equation}
where the sum runs over the relevant non-self-intersecting geodesics $\gamma$ and $\left\langle\cdot\right\rangle$ denotes the gravitational path integral summing over hyperbolic surfaces with any topology.  An important feature of this prescription is that the $\Delta\to0$ limit probes the length while suppressing the backreaction associated with matter insertion. The corresponding Euclidean expression can subsequently be continued to Lorentzian time.

For the case of JT gravity with EoW brane (but unequipped with radiation degrees of freedom), in \cite{Alishahiha:2022kzc} the same procedure as above was applied  to extract the value of the geodesic length but now  from the {\em boundary-to-brane} two-point function of a probe matter field $O$, with scaling dimension $\Delta$ as indicated in eq. \eqref{proposal-geodesic-unequipped}.

 The two-sided correlation function \eqref{proposal-geodesic-unequipped} is obtained by analytic continuation from the corresponding Euclidean correlator, computed by summing over all non self-intersecting geodesics connecting the two points, over all surfaces of arbitrary genus.
 \subsection{Averaging and quenched free energy}
The natural extension of the proposal \eqref{proposal-geodesic-unequipped} to the case of an equipped EoW brane, necessary for mimicking an evaporating black hole, involves considering the ensemble averaged two-point function over the random brane data $\{C_{i,a}\}$, namely
\begin{equation}
   \braket{O_{1}O_{2}}\to \overline{\braket{O_{1}O_{2}}}\,,
\end{equation}
where  $\overline{(\dots)}$ denotes $C$-averaged quantities. The $C$-ensemble average of an object $X$ is defined as, 
\begin{equation}
    \overline{X}=\int dCdC^{\dagger}P[C,C^{\dagger}]X, \quad\quad\quad\quad\quad P[C,C^{\dagger}]\equiv \exp\left({-\text{Tr}(C^{\dagger}C)}\right)\,,
\end{equation}
where $P[C,C^{\dagger}]$ is the average distribution, which is Gaussian by construction.

The un-averaged partition function for gravity plus any matter fields is schematically,
\begin{equation}
    Z=\int \mathcal{D}\zeta \;e^{-I_{E}}\,,\qquad \mathcal{D}\zeta=\frac{\mathcal{D}g\mathcal{D}\phi}{\text{diffeos}}
\end{equation}
To study correlation functions of operators $O_{i}$,\footnote{The operator representations act non-trivially on the states \eqref{psi-random}.} we need to introduce local sources $J^{i}$ for the operators and work with the generating functional,
\begin{equation}
    Z[J^{i}]=\int \mathcal{D}\zeta \;e^{-I_{E}+\sum_{i}J^{i}O_{i}}.
\end{equation}
To obtain the connected correlators, we need its logarithm, the ``free energy",
\begin{equation}
    F[J^{i}]=\log Z[J^{i}]\,,
\end{equation}
so $n$-point correlation functions can be computed as usual,
\begin{equation}
    \braket{O_{1}\cdots O_{m}}_{\text{conn.}}=\frac{\delta F[J^{i}]}{\delta J^{1}\cdots \delta J^{m}}\Bigg|_{J^{i}=0}\,.
\end{equation}
Crucially, in a theory of gravity where wormholes may be present, we need the {\em quenched free energy} since $\overline{\log Z}\neq \log \overline Z$
\cite{Engelhardt:2020qpv} obtained by averaging over the random variables $\{C_{i,a}\}$. This leads us to implement the replica trick for the quenched free energy,
\begin{equation}
    \overline{F[J^{i}]}=\overline{\log{Z[J^{i}]}}=\lim_{n\to 0}\frac{\overline{Z[J^{i}]^{n}}-1}{n}=\lim_{n\to 0}\frac{\partial\overline{Z[J^{i}]^{n}}}{\partial n}\,.
\end{equation}
Therefore we first consider $n$ copies of the theory to compute  $Z[J^{i}]^{n}$ and then perform the ensemble average, finally taking the $n\to 0 $ limit. Schematically then,  we need to evaluate the ensemble averaged $n$-replicated partition function: 
\begin{align}
    \overline{Z[J^{i}]^{n}}&=\int dCdC^{\dagger}P[C,C^{\dagger}]Z^{n}[J^{i}]\\\nonumber
    &=\int dCdC^{\dagger}P[C,C^{\dagger}]\int \prod_{A=1}^{n}\mathcal{D}\zeta_{A}e^{-\sum_{A}S_{E,A}+\sum_{A,i}J^{i}O_{A,i}}\\\nonumber
    &=\int \prod_{A=1}^{n}\mathcal{D}\zeta_{A}\;e^{-S_{E,\text{replica}}+\sum_{A,i}J^{i}O_{i,A}}\,,
\end{align}
where in the last line we have formally performed the integration over $C, C^{\dagger}$, replacing the bulk gravity plus matter action on the replicated geometry, to be viewed as a sum over saddle points.
We can finally write
\begin{align}
    \overline{\braket{O_{1}\cdots O_{m}}_{\text{conn.}}}&=\frac{\delta \overline{F[J_{i}]}}{\delta J^{1}\cdots \delta J^{m}}\Bigg|_{J^{i}=0}\label{quenchedcorr}\\\nonumber
    &=\lim_{n\to 0}\frac{\partial}{\partial n}\big\langle\sum_{A=1}^{n}O_{1,A}\cdots \sum_{B=1}^{n}O_{m,B}\big\rangle_{n}\,\overline{Z[0]^{n}}\,,
\end{align}
where, on the right hand side, correlation functions are evaluated in the $n$-replica manifold and include both connected and disconnected contributions. The factor of $\overline{Z[0]^{n}}$ appears as we are evaluating un-normalised correlators in the replica theory.

Our interest is in the two-point function, with one insertion on the EoW brane and the second on the AdS-boundary:
\begin{equation}
    \braket{O_{1}O_{2}}=e^{-\Delta L_{12}}\,.
\end{equation}
The expectation value of the length operator after the ensemble averaging is, 
\begin{equation}\label{replica_result}
    \overline{L_{12}}=-\lim_{\Delta\to 0}\frac{\partial \overline{e^{-\Delta L_{12}}}}{\partial \Delta}=-\lim_{\Delta\to 0}\frac{\partial}{\partial \Delta}\Biggl[\lim_{n\to 0}\frac{\partial}{\partial n}\big\langle\sum_{A=1}^{n}O_{1,A} \sum_{B=1}^{n}O_{2,B}\big\rangle_{n}\overline{Z[0]^{n}}\Biggr]\,,
\end{equation}
where we will assume that we can interchange the $\Delta\to 0$ and $n\to 0$ limits.

Below we will  compute this geodesic length between the EoW brane and the asymptotic boundary, and given that the black hole interior volume is a candidate for the quantum complexity, the calculation can be viewed as  yielding the complexity evolution for an evaporating black hole.  The result will include non-perturbative contributions due to both  spacetime wormholes and replica wormholes, accounting for  the ensemble average over both the Hamiltonian and the EoW brane states respectively.

\section{Canonical quantisation of JT gravity with EoW brane}
\label{sec4}
In this section we review aspects of canonical quantisation of JT gravity with and without EoW branes, gathering the basic ingredients needed for our calculation.

The canonical quantisation of two-sided JT gravity \cite{Harlow:2018tqv} is in terms of a  two-dimensional phase space with canonical variable the  renormalised geodesic distance $\ell$ between the two endpoints of a time slice, and its conjugate momentum $P$. The Hamiltonian for the system,
\begin{equation}\label{Hamiltonian-pureJT}
    H=2\left(\frac{P^{2}}{4}+e^{-\ell}\right)
 \end{equation}
is that of a non-relativistic particle in an exponential potential. The Hamiltonian is diagonalised by scattering states $\ket{E}$ for $E>0$, with wavefunctions given by modified Bessel functions,
\begin{equation}
    \braket{\ell|E}=4K_{i\sqrt{8E}}(4e^{-\ell/2})=4K_{2is}(4e^{-\ell/2})\,,
\end{equation} 
where  $s^{2}=2E$, and below it will be convenient to work in these energy units.

Let us recall the construction of the two-sided Hartle-Hawking wavefunction in the geodesic basis. Fixing a geodesic of renormalised length $\ell$ across the Euclidean disc and assigning the remaining asymptotic AdS$_2$ boundary length $\beta$, the path integral over geometries of disc topology yields, 
\begin{equation}\label{HH-JT}
    \Phi_{D}(\beta,\ell)=4e^{\frac{S_{0}}{2}}\int_{0}^{\infty}ds\, e^{-\frac{\beta s^{2}}{2}}\rho_{D}(s)K_{2is}(4e^{-\ell/2})\,.
\end{equation}
Here $\rho_{D}(s)$ is the disc density of states, which is given by \cite{Yang:2018gdb, Stanford:2017thb, Kitaev:2018wpr},
\begin{equation}
    \rho_{D}(s)=\frac{s}{2\pi^2}\sinh(2\pi s)\,.
\end{equation}
Upon taking two half discs, each with AdS boundary length $\beta/2$, and sewing them along their geodesic $\ell$, we obtain the JT gravity disc partition function,
\begin{equation}\label{partition-function-pureJT}
    Z_{D}(\beta)=\int_{-\infty}^{+\infty}d\ell\;\Phi_{D}\left(\tfrac{\beta}{2},\ell\right)\Phi_{D}\left(\tfrac{\beta}{2},\ell\right)=e^{S_{0}}\int_{0}^{+\infty}ds\; e^{-\frac{\beta s^{2}}{2}}\rho_{D}(s)\,.
\end{equation}

\subsection{Quantising with an EoW brane}
The canonical quantization of JT gravity in the presence of an EoW brane has been studied in \cite{Gao:2021uro}. The configuration space is parametrised by the  renormalised boundary-to-brane geodesic distance $L$ as the configuration-space coordinate. Canonical quantisation then gives a Hilbert space of square-integrable wavefunctions $\psi(L)$.
The Hamiltonian in this case is that of a particle in a Morse potential, 
\begin{equation} \label{Hamiltonian-EoW}
H=2\left(\frac{P^2}{4}+\mu e^{-L}+e^{-2L}\right)\,.
\end{equation}
 For  $\mu=0$, the Hamiltonian reduces to that of two-sided JT gravity \eqref{Hamiltonian-pureJT} via the identification $\ell=2L$. This is consistent with having a tensionless EoW brane, cutting the two-sided geometry in two identical halves. We restrict below to $\mu>0$. For negative tension the Morse potential develops a negative minimum and supports a discrete bound-state sector, whose geometrical interpretation is a naked EoW brane configuration rather than the black-hole geometries relevant here \cite{Gao:2021uro}. 

Setting $z=4e^{-L}$, the eigenstates are given by Whittaker functions of the second kind
\begin{equation}\label{wavefunction-EoW}
\braket{L|s}=\sqrt{f_\mu(s)}\; \frac{W_{-\mu,is}(z)}{\sqrt{z}}\,,\qquad f_\mu(s)=2
\gamma_\mu(s)\rho_{D}(s)\,,
\end{equation}
where we have defined 
\begin{equation} \label{EoWfunction}
\gamma_\mu(s)=\left|\Gamma
\left(\tfrac{1}{2}+\mu+is\right)\right|^2\,.
\end{equation}
The normalisation of the wavefunction \eqref{wavefunction-EoW} follows from the orthogonality property of the Whittaker functions,
\begin{equation}\label{OW}
\int_0^\infty \frac{dz}{z^2} \;W_{-\mu,is}(z)\; W_{-\mu, is'}(z)=\frac{1}{f_\mu(s)}\delta(s-s')\,.
\end{equation}

We can now construct the Hartle-Hawking wavefunction $\Psi_D(\beta,L)$ for the one-sided geometry with an EoW brane \cite{Alishahiha:2022kzc}, prior to introducing its internal states. It is defined by the Euclidean disc path integral with the geodesic separation $L$ between the EoW brane and the asymptotic AdS boundary held fixed, while the latter has renormalised length $\beta$:
\begin{equation}\label{HH-EoW}
    \Psi_{D}(\beta,L)=\sqrt{2}e^{\frac{S_{0}}{2}}\int_{0}^{\infty}ds\, e^{-\frac{\beta s^{2}}{2}}\gamma_{\mu}(s)\rho_{D}(s)\frac{W_{-\mu,is}(z)}{\sqrt{z}}\,.
\end{equation}
The disc partition function is computed as before by sewing two such half-discs,
\begin{equation}\label{partition-function-EoW}
    Z_{D,\mu}(\beta)=\int_{0}^{\infty}\frac{dz}{z}\Psi_{D}\left(\tfrac{\beta}{2},L\right)\Psi_{D}\left(\tfrac{\beta}{2},L\right)=e^{S_{0}}\int_{0}^{\infty}ds\, e^{-\frac{\beta s^{2}}{2}}\gamma_{\mu}(s)\rho_{D}(s)
\end{equation}
Including the EoW brane changes the disc density of states,
\be
\rho_{D}(s) \to \rho_{D}(s) \left|\Gamma
\left(\tfrac{1}{2}+\mu+is\right)\right|^2\,,
\ee
modifying the large-$s$ growth of the density of states.
\begin{figure}[ht]
\centering
\begin{tikzpicture}

    \draw[thick] (0,0) arc[start angle=-180,end angle=0,radius=1.8cm] node[midway, below] {$\beta$};

    \draw[red, line width=1pt]
        (0,0) .. controls (1.8,0.2) .. (3.6,0) node[midway, above] {$\ell$};

    \draw[thick] (9,0) arc[start angle=0,end angle=-120,radius=1.8cm] node[midway, below] {$\beta$};

    \draw[blue, line width=1pt]
        ({7.2+1.8*cos(120)}, {-1.8*sin(120)}) .. controls (6.6,-0.9) .. (6.8,0) node[midway, left] {$\mu$};
    \draw[red, line width=1pt]
        (6.8,0).. controls (7.9,0.2) .. (9,0) node[midway, above] {$L$};
\end{tikzpicture}
\caption{Hartle-Hawking wavefunctions, $\Phi_D(\beta, \ell)$ for   two-sided JT black hole (left) and $\Psi_D(\beta, L)$ for the single-sided JT black hole with an EoW brane (right).}
\end{figure}
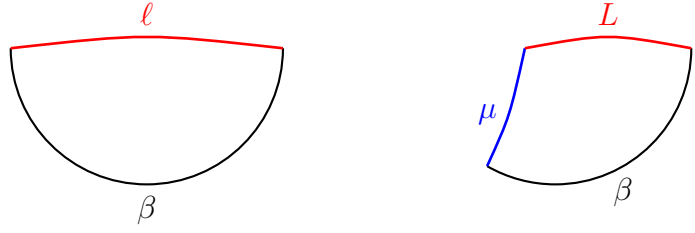

\section{Ensemble averaged geodesic length}
\label{sec5}
To compute the geodesic length employing a quenched averaging of $\log Z$, we need to gather the ingredients of the associated replica calculation. Key amongst these is the Hartle-Hawking wave function for the $n$-boundary geometry to which we first turn our attention.
\subsection{$n$-boundary wave function}
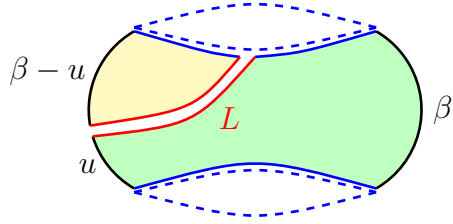
\begin{figure}[H]
\centering
\begin{tikzpicture}

     \fill[yellow!30]
    (8,0) ++(120:1.2cm)
    arc[start angle=120, end angle=190, radius=1.2cm]
    .. controls (8.2,0) .. (8.8,0.7)
    .. controls (8.4,0.75) .. ({8+1.2*cos(120)}, {1.2*sin(120)})
    -- cycle;

    \fill[green!25]
    (8,0) ++(197:1.2cm)
    arc[start angle=197, end angle=240, radius=1.2cm]
    .. controls (9,-0.6) .. ({10+1.2*cos(60)}, {-1.2*sin(60)})
    arc[start angle=-60, end angle=60, radius=1.2cm]
    .. controls (9.6,0.75) .. (9,0.7)
    .. controls (8.2,-0.2) .. ({8+1.2*cos(197)}, {1.2*sin(197)})
    -- cycle;
    
    \draw[black, line width=1pt] (8,0) ++(120:1.2cm) arc [start angle=120, end angle=190, radius=1.2cm] node[midway, left] {$\beta-u$};

    \draw[black, line width=1pt] (8,0) ++(197:1.2cm) arc [start angle=197, end angle=240, radius=1.2cm] node[midway, left] {$u$};

    \draw[black, line width=1pt] (10,0) ++(-60:1.2cm) arc [start angle=-60, end angle=60, radius=1.2cm] node[midway, right] {$\beta$};

    \draw[blue, line width=1pt] ({8+1.2*cos(120)}, {1.2*sin(120)} ).. controls (8.4,0.75)..(8.8,0.7);

    \draw[blue, line width=1pt] (9, 0.7).. controls (9.6,0.75).. ({10+1.2*cos(60)}, {1.2*sin(60)});

    \draw[red, line width=1pt] ({8+1.2*cos(190)}, {1.2*sin(190)}) .. controls (8.2,0).. (8.8,0.7);

    \draw[red, line width=1pt] ({8+1.2*cos(197)}, {1.2*sin(197)}) .. controls (8.2,-0.2).. (9,0.7) node[midway, right] {$\;\;L$};




    \draw[blue, dashed, line width=1pt] ({8+1.2*cos(120)}, {0.05+1.2*sin(120)} ).. controls (8+1,0.7)..({8+2+1.2*cos(60)}, {0.05+1.2*sin(60)});

    \draw[blue, dashed, line width=1pt] ({8+1.2*cos(120)}, {0.05+1.2*sin(120)} ).. controls (8+1,1.5)..({8+2+1.2*cos(60)}, {0.05+1.2*sin(60)});

    \draw[blue, dashed, line width=1pt] ({8+1.2*cos(120)}, {-0.05-1.2*sin(120)} ).. controls (8+1,-0.7)..({8+2+1.2*cos(60)}, {-0.05-1.2*sin(60)});

    \draw[blue, dashed, line width=1pt] ({8+1.2*cos(120)}, {-0.05-1.2*sin(120)} ).. controls (8+1,-1.5)..({8+2+1.2*cos(60)}, {-0.05-1.2*sin(60)});

    \draw[blue, line width=1pt] ({8+1.2*cos(120)}, {-1.2*sin(120)} ).. controls (9,-0.6)..({10+1.2*cos(60)}, {-1.2*sin(60)});




\end{tikzpicture}
\caption{\footnotesize Pictorial representation of the two disc-like geometries associated to the generalized Hartle-Hawking wavefunctions $\Psi_{2}(\beta+u,L)$ (shaded in green) and $\Psi_{1}(\beta-u,L)$ (shaded in yellow), whose gluing gives the path integral $Z_{2}$. From now on, dashed blue lines stand for identification of EoW-brane flavour indices.}
\label{Z2_decomposition}
\end{figure}

The gravitational path integral on a completely connected disc-like wormhole geometry bounded by $n$ asymptotic boundaries and $n$ EoW branes is given by \cite{Penington:2019kki},
\begin{equation}\label{Zn}
    Z_{n}=e^{S_{0}}\int_{0}^{\infty} ds\,\rho_{D}(s) y(s)^{n}, \quad\quad\quad y(s)=e^{-\frac{\beta s^{2}}{2}}\gamma_{\mu}(s)\,.
\end{equation}
Having $n$ boundaries, each of length $\beta$, has the effect of rescaling $\beta$ by a factor of $n$, and the introduction of each EoW brane with tension $\mu$ is accompanied by a factor of $\gamma_\mu(s)$ in the effective density of states. The corresponding generalised Hartle-Hawking state wave function is,
\begin{equation}\label{generalized, HH}
    \Psi_{n}(x,L)=\sqrt{2}e^{\frac{S_{0}}{2}}\int_{0}^{+\infty} ds\,\rho_{D}(s)\,e^{-\frac{xs^2}{2}}\gamma_{\mu}(s)^{n}\,\frac{W_{-\mu,\,is}(z)}{\sqrt{z}}\,.
\end{equation}
This is the outcome of a gravitational path integral over  geometries with $n$  AdS boundaries of total length $x$, $n$ EoW branes, and one geodesic of length $L$ stretching from an EoW brane to an AdS boundary. 
In particular, it can be  verified that the $n=2$ partition function $Z_{2}$ is obtained by gluing two generalised Hartle-Hawking states, as depicted in Figure \ref{Z2_decomposition},
\begin{equation}\label{Z2, HH}
    Z_{2}=\int_{0}^{+\infty} \frac{dz}{z}\Psi_{1}(\beta-u,L)\Psi_{2}(\beta+u,L)\,.
\end{equation}
 The equality \eqref{Z2, HH} follows from the normalization property \eqref{OW} of the Whittaker functions.
 
The standard procedure for deriving \eqref{generalized, HH} is to first consider the gravitational path integral over the disc with $2n+1$ geodesic boundaries of fixed lengths \cite{Penington:2019kki}
\begin{equation}
    I_{2n+1}(\ell_{1},\dots,\ell_{2n}, \ell_{2n+1})=2^{2n+1}\int_{0}^{+\infty}ds\, \rho_{D}(s) \prod_{j=1}^{2n+1}K_{2is}(4e^{-\ell_{j}/2})
    \,.
\end{equation}
This object assumes that each geodesic stretches between two asymptotic AdS$_2$ boundaries. If we wish to  replace one of the geodesics with a geodesic  stretching between an asymptotic boundary and an EoW brane,  this can be achieved by replacing the corresponding Bessel function with a Whittaker function,
\begin{equation}
    I_{2n+1}(\ell_{1},\dots,\ell_{2n},L)\to 2^{2n-1/2}\int_{0}^{+\infty}ds\, \rho_{D}(s) \prod_{j=1}^{2n}K_{2is}(4e^{-\ell_{j}/2})
    W_{-\mu,\,is}(4e^{-L/2})\,.
\end{equation}
To obtain the generalised Hartle-Hawking state $\Psi_{n}$ with $L$ fixed, we must then integrate over the remaining $2n$ geodesic lengths with a suitable wavefunction for each geodesic, which is either $2^{(1+2\mu)}e^{-(\mu+\frac12)\ell}$ for an EoW brane\footnote{This wavefunction differs from $e^{-\mu \ell}$, derived in Appendix D of \cite{Penington:2019kki}, since we chose a different set of brane states $\ket{\psi_{i}(\beta)}$ \eqref{psi-random} to make contact with the canonical quantization of JT with EoW brane carried out in \cite{Gao:2021uro}. Despite this basis difference, the results of this paper will be independent of it, since we will work either in the microcanonical ensemble or in the large brane tension limit.} or the disc wavefunction \eqref{HH-JT} $\Phi_D(\beta, \ell)$ to account for an asymptotic AdS$_2$ boundary.\footnote{Here we need to be careful to remove the factor of $e^{-S_0/2}$ in \eqref{HH-JT}.}
This procedure leads to the generalised Hartle-Hawking state in the case with $n$ EoW branes, and $n$ AdS$_2$ boundaries. Using standard integrals involving the Bessel functions, we find,
\begin{align}\label{HH-general}
    &\Psi_{n}(n\beta,L)=\\
    &\qquad e^{\frac{S_{0}}{2}}2^{(1+2\mu)n}\int \prod_{i=1}^{2n}{d\ell_{i}}\;e^{L/2}\,
    I_{2n+1}(\ell_{1},\dots, \ell_{2n},L)\prod_{j=1}^{n}\left(e^{-\frac{S_0}{2}}\Phi_D(\beta,\ell_{2j-1})e^{-(\mu+\frac12) \ell_{2j}}\right)\notag\\\nonumber
    &\qquad=\sqrt{2}\,e^{\frac{S_{0}}{2}}\int_{0}^{+\infty}ds\,\rho_{D}(s)\,e^{-\frac{n\beta s^{2}}{2}}\gamma_{\mu}(s)^{n}\frac{W_{-\mu,\,is}(z)}{\sqrt{z}}\,.
\end{align}
Integration over the $2n$ geodesics requires $\ell_j$-dependent measure  factors \cite{Yang:2018gdb,Penington:2019kki} which have already been accounted for in arriving at the expression above. Note also that the final result scales with $e^{S_0/2}$ as expected for the disc wave function; factors of $e^{-S_0/2}$ in the first line of  \eqref{HH-general} are needed to strip off the corresponding normalisation in $\Phi_D$ as defined in \eqref{HH-JT}\,.

\subsection{Quenched average: replica computation}
Exploiting the wavefunctions derived above, we want to find a closed form expression for the expectation value of the interior length $L$ in a general replica geometry,
\begin{equation} \label{summed-replica-corr}
    \braket{L(t)}_{n}\equiv-\frac1k\cdot\lim_{\Delta\to 0}\frac{\partial}{\partial \Delta}\left(\big\langle\sum_{A=1}^{n}O_{1,A} \sum_{B=1}^{n}O_{2,B}\big\rangle\,\overline{Z^{n}}\right)\,.
\end{equation}
Eventually we will take the $n\to 0$ limit and arrive at the expectation value of the geodesic length,
\begin{equation}
    \overline{\braket{L(t)}}=\lim_{n\to 0}\frac{\partial}{\partial n}\braket{L(t)}_{n}\,.
\end{equation}
We draw attention to the factor of $1/k$ that appears in \eqref{summed-replica-corr}. This is an important ingredient in what follows and has implications for the physics we infer. It is required in order to yield the correctly normalised $\braket {O_1 O_2}$ correlator. Since the operators $O_{1,2}$ do not carry any brane/flavour indices, their normalised correlator in the absence of replica wormholes must have no $k$-dependence, thus the extra factor of $1/k$ is required to ensure that the completely disconnected contributions to our replica correlator do not have any $k$ dependence. 

\paragraph{Example $n=2$:}It is instructive to first consider the $n=2$ case to unpack the calculation of 
$\braket{L(t)}_{n}$ for generic integer $n$. The unnormalized $n=2$ replica correlation functions inside the $\Delta$-derivative in \eqref{summed-replica-corr} can be written as
\begin{align}
    \big\langle\sum_{A=1}^{2}O_{1,A} \sum_{B=1}^{2}O_{2,B}\big\rangle\ \overline{Z^{2}} =&2kZ_1\int_{0}^{\infty}\frac{dz}{z}\Psi_{1}(\beta-u,L)\Psi_{1}(u,L)\left(\frac{z}{4}\right)^{\Delta} +\label{OOn=2}\\
    + &2k^2\int_{0}^{\infty}\frac{dz}{z}\Psi_{2}(2\beta-u,L)\Psi_{1}(u,L)\left(\frac{z}{4}\right)^{\Delta}\,+\notag\\
    +&2k^2\int_{0}^{\infty}\frac{dz}{z}\Psi_{2}(\beta+u,L)\Psi_{1}(\beta-u,L)\left(\frac{z}{4}\right)^{\Delta}\,.\notag
\end{align}
The first term accounts for two disconnected copies of the single-disc geometry, with a geodesic constructed in one of them connecting the EoW brane to the AdS boundary.  The second and third terms represent geodesic contributions built in the wormhole connected replica geometries (Figure \ref{Z2_decomposition} and its mirror image). Each contribution is weighted by an appropriate power of $k$, which counts the number of EoW brane index loops. 

Taking the $\Delta$-derivative and following the steps detailed  in appendix \ref{review-IMS}, we find,
\bea\label{braketLn}
   \braket{L_k(t)}_{n=2}= &&L_{k,{\rm div}}\big|_{n=2}\\\nonumber
  - &&\frac{2\pi e^{2S_0}}{3 k}\int_{E_*}^\infty dE\,
\,
{\cal I}(n=2)\,\sqrt{2E}\rho_D(E)^2\, 
\left(1-\frac{t}{2 \pi  \rho_D(E)  e^{S_0}}\right)^3\,,
\eea
where we have introduced a subscript label $k$ for the geodesic length in the theory with $k$ EoW brane states. A noteworthy point which will be important subsequently, is that the divergent contributions from energy integrals depend nontrivially on $k$ and $n$.
As in \cite{Alishahiha:2022kzc,Iliesiu:2021ari}, central to the derivation of this intermediate result  is the non-perturbative spectral two-point function in JT gravity \eqref{eq:sinekernel} which includes spacetime wormhole effects joining the two halves of the disc geometries sliced by the geodesic.  Crucially, the {\it new element} in the present context is the kernel ${\cal I}(n)$ in the energy integral above, which also includes the effect of  replica wormholes on the disc geometries. We describe this in some detail below, first breaking up the individual contributions to ${\cal I}(n=2)$ and explaining their combinatorial origin via the generalisation for arbitrary $n$,
\begin{equation}\label{I2}
    {\cal I}(n=2)=e^{-\beta E}\gamma_{\mu}(E)\,k\times 1\times 1\times \binom{2}{1}{\color{orange}Z_1}+e^{-2\beta E}\gamma_{\mu}(E)^{2}\,k^{2}\times 2\times 2\times \binom{2}{2}\,.
\end{equation}
\begin{figure}[H]
\centering
\begin{tikzpicture}[baseline]
     \draw[black, line width=1pt] (0,0) ++(60:1.2cm) arc [start angle=60, end angle=185, radius=1.2cm] node[midway, left] {$\beta-u\;\;$};
     \draw[black, line width=1pt] (0,0) ++(192:1.2cm) arc [start angle=192, end angle=300, radius=1.2cm] node[midway, left] {$u\;\;\;$};

    \draw[blue, line width=1pt] ({1.2*cos(60)}, {1.2*sin(60)} ).. controls (0.4,0.55)..(0.3,0.1);
    \draw[blue, line width=1pt] (0.3,0).. controls (0.4,-0.55)..( {1.2*cos(60)}, { -1.2*sin(60)});

    \draw[blue, dashed, line width=1pt] ({1.2*cos(57)}, {1.2*sin(57)} ).. controls (0.3,0)..( {1.2*cos(57)}, { -1.2*sin(57)});

    \draw[red, line width=1pt] ({1.2*cos(185)}, {1.2*sin(185)}).. controls (-0.3, 0.2)..(0.3,0.1);
    \draw[red, line width=1pt] ({1.2*cos(192)}, {1.2*sin(192)}).. controls (-0.3, 0.1)..(0.3,0);

    \fill[orange]
    (3,0) ++(-120:1.2cm)
      arc[start angle=-120, end angle=120, radius=1.2cm]
      .. controls (2.8,0) ..
        ({3+1.2*cos(-120)}, {1.2*sin(-120)})
    -- cycle;
    
    \draw[black, line width=1pt] (3,0) ++(-120:1.2cm) arc [start angle=-120, end angle=120, radius=1.2cm];
        \draw[blue, line width=1pt] ({3+1.2*cos(120)}, {1.2*sin(120)} ).. controls (2.8,0)..({3+1.2*cos(120)}, {-1.2*sin(120)});
        \draw[blue, dashed, line width=1pt] ({3+1.2*cos(123)}, {1.2*sin(123)} ).. controls (2.7,0)..({3+1.2*cos(123)}, {-1.2*sin(123)});


        \draw[blue, dashed, line width=1pt] ({1.2*cos(57)}, {1.2*sin(57)} ).. controls (1.5,1.3)..( {3+1.2*cos(123)}, {1.2*sin(123)});
        \draw[blue, dashed, line width=1pt] ({1.2*cos(57)}, {-1.2*sin(57)} ).. controls (1.5,-1.3)..( {3+1.2*cos(123)}, {-1.2*sin(123)});
        
    \draw[black, line width=1pt] (8,0) ++(120:1.2cm) arc [start angle=120, end angle=190, radius=1.2cm] node[midway, left] {$\beta-u$};

    \draw[black, line width=1pt] (8,0) ++(197:1.2cm) arc [start angle=197, end angle=240, radius=1.2cm] node[midway, left] {$u$};

    \draw[black, line width=1pt] (10,0) ++(-60:1.2cm) arc [start angle=-60, end angle=60, radius=1.2cm] node[midway, right] {$\beta$};

    \draw[blue, line width=1pt] ({8+1.2*cos(120)}, {1.2*sin(120)} ).. controls (8.4,0.75)..(8.8,0.7);

    \draw[blue, line width=1pt] (9, 0.7).. controls (9.6,0.75).. ({10+1.2*cos(60)}, {1.2*sin(60)});

    \draw[red, line width=1pt] ({8+1.2*cos(190)}, {1.2*sin(190)}) .. controls (8.2,0).. (8.8,0.7);

    \draw[red, line width=1pt] ({8+1.2*cos(197)}, {1.2*sin(197)}) .. controls (8.2,-0.2).. (9,0.7);

    \draw[blue, dashed, line width=1pt] ({8+1.2*cos(120)}, {0.05+1.2*sin(120)} ).. controls (8+1,0.7)..({8+2+1.2*cos(60)}, {0.05+1.2*sin(60)});

    \draw[blue, dashed, line width=1pt] ({8+1.2*cos(120)}, {0.05+1.2*sin(120)} ).. controls (8+1,1.5)..({8+2+1.2*cos(60)}, {0.05+1.2*sin(60)});

    \draw[blue, dashed, line width=1pt] ({8+1.2*cos(120)}, {-0.05-1.2*sin(120)} ).. controls (8+1,-0.7)..({8+2+1.2*cos(60)}, {-0.05-1.2*sin(60)});

    \draw[blue, dashed, line width=1pt] ({8+1.2*cos(120)}, {-0.05-1.2*sin(120)} ).. controls (8+1,-1.5)..({8+2+1.2*cos(60)}, {-0.05-1.2*sin(60)});
    



    \draw[blue, line width=1pt] ({8+1.2*cos(120)}, {-1.2*sin(120)} ).. controls (9,-0.6)..({10+1.2*cos(60)}, {-1.2*sin(60)});



\end{tikzpicture}
\caption{\footnotesize Topologically distinct contributions to ${\cal I}(n=2)$.}
\label{slicedgeometry}
\end{figure}
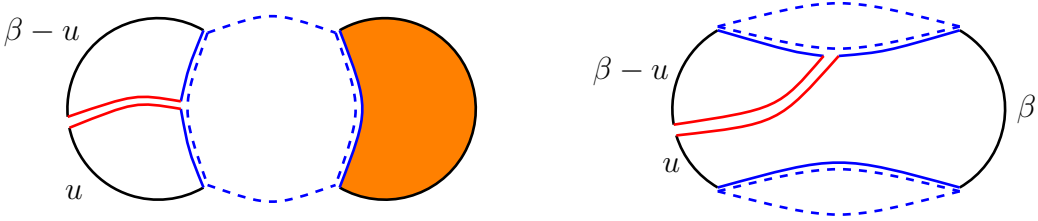
\paragraph{General $n$ prescription:} For  general $n$, we can write,
\be
{\cal I}(n)= \sum_{j=1}^{n} {\cal I}_j(n)\,,
\ee
where the $\{{\cal I}_j\}$ are schematically,
\bea
&&{\cal I}_j(n) =\\\nonumber
&&\underbrace{\underbrace{e^{-j\beta E}}_{\text{boundary weight}}\times\underbrace{\gamma_{\mu}(E)^{j}}_{\text{brane weight}} \times \underbrace{k^{j}}_{\text{index loops}} \times \underbrace{j}_{\text{\#geodesics}} \times \underbrace{j}_{\text{\#branes}} \times \underbrace{\binom{n}{j}}_{\text{topology}}}_{\text{geometry sliced by the geodesic}} \times \underbrace{(\dots)}_{\text{un-sliced pieces}}\,.
\eea
Each ${\cal I}_j(n)$ represents a specific topology of the $n$-replica geometry with a $j$-boundary\footnote{Here we refer to the number of AdS boundaries of the geometry, which also coincides with the number of EoW boundaries.} connected component in  which the geodesic from an EoW brane resides, accompanied by the remaining components with $(n-j)$ AdS boundaries which can include both connected and disconnected pieces. Since the geodesic slices through the $j$-boundary connected component we will call this the ``sliced" geometry (see Figure \ref{slicedgeometry}). This is accompanied by a combinatorial factor $\binom{n}{j}$ from picking $j$ out of $n$ boundaries.  Each sliced geometry has $j$-boundaries of total length $j\beta$ which is reflected in the Boltzmann factor $e^{-j\beta E}$ or ``boundary weight". Since there are also $j$ EoW branes, we get a ``brane weighting" factor $\gamma_\mu(E)^j$. Every brane boundary is accompanied by an index loop which sums over the $k$ brane states yielding an additional factor of $k^j$. Finally, within a given sliced geometry, the  geodesic can begin at any one of $j$ possible EoW branes and end on any one of $j$ possible AdS boundaries. It is not {\em a priori} obvious that these different endpoint choices should yield identical contributions, nonetheless they do when the timescales in question are large compared to $\beta$, the thermal time, as clarified in a moment.

The “un-sliced" contribution to ${\cal I}_j$ is from the  path integral over all possible geometries which can be constructed with the remaining $n-j$ asymptotic boundary/EoW brane pairs, along with their respective topology-specific binomial factors and  brane index loop factors.

We now make some remarks on the approximations made in getting to  eq. \eqref{braketLn} and the kernel \eqref{I2}. Firstly, in the energy integrals arising from the two-point functions, we have worked to leading nontrivial order in the coincidence limit for the spectral two-point functions $\langle \rho(E_1)\rho(E_2)\rangle$, which means that we  have worked in the small $\omega$ limit ($\omega = E_1 - E_2$) and set contributions $\sim e^{\omega \beta}$ to unity (so $\omega\beta$ is small and the time scales of our interest are much larger than thermal time $\beta$). Such terms enter into the boundary weights of 
generalized Hartle-Hawking wavefunctions \eqref{generalized, HH}, and yield additional constant contributions to the $\omega$-integrals in  \eqref{finalfreq} . We have checked the consistency of this truncation in appendix \ref{AppendixE1}.

Second, we do not consider handles\footnote{Apart from the spacetime wormholes connecting the two pieces involved in the geodesic slicing, which we have been accounting for by implementing \eqref{replacement}.} which increase the genus and are suppressed by powers of $e^{-2S_{0}}$, or brane crossing terms which are suppressed by powers of $k^{-2}$. These would introduce subleading corrections, as was also the case for the derivation of the Page curve within this evaporating black hole model \cite{Penington:2019kki, Iizuka:2024njd}. We also  do not include EoW brane loops, since all the brane geodesics must begin and end on an asymptotic AdS boundary.
Finally, one may argue  that we should also include geodesics not cutting through a disc geometry, rather joining local operator insertions in disconnected geometries. It is unclear how to interpret this physically, since the resulting geodesics do not effectively stretch through any bulk geometry, but it can be checked using the boundary particle formalism that such contributions yield time-independent divergent contributions, which can be subtracted away by an appropriate renormalisation prescription.

\paragraph{Example $n=3$:} We will generalize the structure leading to eq.\,\eqref{braketLn} to arbitrary positive integer $n$ and find a closed form for the kernel ${\cal I}(n)$. Before turning to this, we glance at $n=3$ as an additional illustrative example:
\be
{\cal I}(3)= {\cal I}_1(3)+{\cal I}_2(3)+{\cal I}_3(3)\,,
\ee
where,
\be
    {\cal I}_3(3)=e^{-3\beta E}\,\gamma_{\mu}(E)^{3}\,k^{3}\times 3\times 3\times \binom{3}{3}\,\label{n=3, L,1}\,,
    \ee
    \be
    {\cal I}_2(3)=e^{-2\beta E}\,\gamma_{\mu}(E)^{2}\,k^{2}\times 2\times 2\times \binom{3}{2}\,{\color{orange}Z_1}\,\label{n=3, L,2}\,,
    \ee
    \be
    {\cal I}_1(3)=e^{-\beta E}\,\gamma_{\mu}(E)\,k\times 1\times 1\times \binom{3}{1}\,({\color{yellow!70!orange}Z_1^2}\,+\,k\,{\color{green!60!black}Z_{2}})\label{n=3, L,3}\,.
\ee
\begin{figure}[ht]
\centering
\begin{tikzpicture}[baseline]
\node at (-3,0) {${\cal I}_3(3)$:};
    \draw[black, line width=1pt] (0,0) ++(125:1.2cm) arc [start angle=125, end angle=152, radius=1.2cm];
     \draw[black, line width=1pt] (0,0) ++(158:1.2cm) arc [start angle=158, end angle=185, radius=1.2cm];
     \draw[black, line width=1pt] (0,0) ++(240:1.2cm) arc [start angle=240, end angle=300, radius=1.2cm];
     \draw[black, line width=1pt] (0,0) ++(-5:1.2cm) arc [start angle=-5, end angle=55, radius=1.2cm];

    \draw[blue, dashed, line width=1pt] (0,0) ++(55:1.2cm) arc [start angle=55, end angle=125, radius=1.2cm];
    \draw[blue, dashed, line width=1pt] (0,0) ++(300:1.2cm) arc [start angle=300, end angle=355, radius=1.2cm];
    \draw[blue, dashed, line width=1pt] (0,0) ++(185:1.2cm) arc [start angle=185, end angle=240, radius=1.2cm];

    \draw[blue, line width=1pt] ({1.2*cos(55)}, {1.2*sin(55)} ).. controls (0.6,0.9)..(0.1, 0.7);
    \draw[blue, line width=1pt] (-0.1, 0.7).. controls (-0.6, 0.9)..({1.2*cos(125)}, {1.2*sin(125)} );

    \draw[blue, dashed, line width=1pt] ({1.2*cos(189)}, {1.2*sin(189)} ).. controls (-0.7,-0.35)..({1.2*cos(236)}, {1.2*sin(236)} );
    \draw[blue, line width=1pt] ({1.2*cos(355)}, {1.2*sin(355)} ).. controls (0.6,-0.3)..({1.2*cos(300)}, {1.2*sin(300)} );
    \draw[blue, dashed, line width=1pt] ({1.2*cos(351)}, {1.2*sin(351)} ).. controls (0.7,-0.35)..({1.2*cos(304)}, {1.2*sin(304)} );
    \draw[blue, line width=1pt] ({1.2*cos(185)}, {1.2*sin(185)} ).. controls (-0.6,-0.3)..({1.2*cos(240)}, {1.2*sin(240)} );
    \draw[blue, dashed, line width=1pt] ({1.2*cos(59)}, {1.2*sin(59)} ).. controls (0, 0.7)..({1.2*cos(121)}, {1.2*sin(121)} );

    \draw[red, line width=1pt] ({1.2*cos(152)}, {1.2*sin(152)} ).. controls (-0.4,0.4)..(-0.1, 0.7);
    \draw[red, line width=1pt] ({1.2*cos(158)}, {1.2*sin(158)} ) .. controls (-0.4,0.3)..(0.1, 0.7);

\begin{scope}[xshift=4.0cm]
    \draw[black, line width=1pt] (0,0) ++(125:1.2cm) arc [start angle=125, end angle=185, radius=1.2cm];
     \draw[black, line width=1pt] (0,0) ++(240:1.2cm) arc [start angle=240, end angle=266, radius=1.2cm];
     \draw[black, line width=1pt] (0,0) ++(274:1.2cm) arc [start angle=274, end angle=300, radius=1.2cm];
     \draw[black, line width=1pt] (0,0) ++(-5:1.2cm) arc [start angle=-5, end angle=55, radius=1.2cm];

     \draw[blue, dashed, line width=1pt] (0,0) ++(55:1.2cm) arc [start angle=55, end angle=125, radius=1.2cm];
    \draw[blue, dashed, line width=1pt] (0,0) ++(300:1.2cm) arc [start angle=300, end angle=355, radius=1.2cm];
    \draw[blue, dashed, line width=1pt] (0,0) ++(185:1.2cm) arc [start angle=185, end angle=240, radius=1.2cm];

    \draw[blue, line width=1pt] ({1.2*cos(55)}, {1.2*sin(55)} ).. controls (0.6,0.9)..(0.1, 0.7);
    \draw[blue, line width=1pt] (-0.1, 0.7).. controls (-0.6, 0.9)..({1.2*cos(125)}, {1.2*sin(125)} );

    \draw[blue, dashed, line width=1pt] ({1.2*cos(189)}, {1.2*sin(189)} ).. controls (-0.7,-0.35)..({1.2*cos(236)}, {1.2*sin(236)} );
    \draw[blue, line width=1pt] ({1.2*cos(355)}, {1.2*sin(355)} ).. controls (0.6,-0.3)..({1.2*cos(300)}, {1.2*sin(300)} );
    \draw[blue, dashed, line width=1pt] ({1.2*cos(351)}, {1.2*sin(351)} ).. controls (0.7,-0.35)..({1.2*cos(304)}, {1.2*sin(304)} );
    \draw[blue, line width=1pt] ({1.2*cos(185)}, {1.2*sin(185)} ).. controls (-0.6,-0.3)..({1.2*cos(240)}, {1.2*sin(240)} );
    \draw[blue, dashed, line width=1pt] ({1.2*cos(59)}, {1.2*sin(59)} ).. controls (0, 0.7)..({1.2*cos(121)}, {1.2*sin(121)} );

    \draw[red, line width=1pt] ({-0.1}, {0.72} )--({-0.1},-1.2);
    \draw[red, line width=1pt] ({0.1}, {0.72} )--({0.1},-1.2) ;
\end{scope}
\end{tikzpicture}

 \vspace{2em}

\begin{tikzpicture}[xshift=-2cm]
\node at (-3.5,0) {${\cal I}_2(3)$:};
\node[text=white] at (4,0) {${\cal I}_2(3)$:};
    \fill[orange]
    (0,0) ++(-5:1.2cm)
      arc[start angle=-5, end angle=55, radius=1.2cm]
      .. controls (0.6,0.3) ..
        ({1.2*cos(-5)}, {1.2*sin(-5)})
    -- cycle;

    \draw[black, line width=1pt] (0,0) ++(125:1.2cm) arc [start angle=125, end angle=152, radius=1.2cm];
     \draw[black, line width=1pt] (0,0) ++(158:1.2cm) arc [start angle=158, end angle=185, radius=1.2cm];
     \draw[black, line width=1pt] (0,0) ++(240:1.2cm) arc [start angle=240, end angle=300, radius=1.2cm];
     \draw[black, line width=1pt] (0,0) ++(-5:1.2cm) arc [start angle=-5, end angle=55, radius=1.2cm];

    \draw[blue, dashed, line width=1pt] (0,0) ++(55:1.2cm) arc [start angle=55, end angle=125, radius=1.2cm];
    \draw[blue, dashed, line width=1pt] (0,0) ++(300:1.2cm) arc [start angle=300, end angle=355, radius=1.2cm];
    \draw[blue, dashed, line width=1pt] (0,0) ++(185:1.2cm) arc [start angle=185, end angle=240, radius=1.2cm];

    \draw[blue, line width=1pt] ({1.2*cos(125)}, {1.2*sin(125)} ).. controls (-0.2,0.6)..(0,0.3);
    \draw[blue, line width=1pt] (0.1,0.2).. controls (0.3,-0.1)..({1.2*cos(300)}, {1.2*sin(300)} );
    \draw[blue, line width=1pt] ({1.2*cos(185)}, {1.2*sin(185)} ).. controls (-0.7,-0.5)..({1.2*cos(240)}, {1.2*sin(240)} );
    \draw[blue, line width=1pt] ({1.2*cos(355)}, {1.2*sin(355)} ).. controls (0.6,0.3)..({1.2*cos(55)}, {1.2*sin(55)} );
    \draw[blue, dashed, line width=1pt] ({1.2*cos(351)}, {1.2*sin(351)} ).. controls (0.55,0.2)..({1.2*cos(59)}, {1.2*sin(59)} );

    \draw[blue, dashed, line width=1pt] ({1.2*cos(121)}, {1.2*sin(121)} ).. controls (0.3,0.18)..({1.2*cos(304)}, {1.2*sin(304)} );
    \draw[blue, dashed, line width=1pt] ({1.2*cos(189)}, {1.2*sin(189)} ).. controls (-0.8,-0.55)..({1.2*cos(236)}, {1.2*sin(236)} );

    \draw[red, line width=1pt] ({1.2*cos(152)}, {1.2*sin(152)} ).. controls (-0.4,0.27)..(0,0.3);
    \draw[red, line width=1pt] ({1.2*cos(158)}, {1.2*sin(158)} ) .. controls (-0.4,0.18)..(0.1,0.2);
\end{tikzpicture}

\vspace{2em}

\begin{tikzpicture}[baseline]
\node at (-3,0) {${\cal I}_1(3)$:};
    \fill[yellow]
    (0,0) ++(-5:1.2cm)
      arc[start angle=-5, end angle=55, radius=1.2cm]
      .. controls (0.5,0.2) ..
        ({1.2*cos(-5)}, {1.2*sin(-5)})
    -- cycle;

    \fill[yellow]
    (0,0) ++(240:1.2cm)
      arc[start angle=240, end angle=300, radius=1.2cm]
      .. controls (0,-0.6) ..
        ({1.2*cos(240)}, {1.2*sin(240)})
    -- cycle;
    
     \draw[black, line width=1pt] (0,0) ++(125:1.2cm) arc [start angle=125, end angle=158, radius=1.2cm];
     \draw[black, line width=1pt] (0,0) ++(164:1.2cm) arc [start angle=164, end angle=185, radius=1.2cm];
     \draw[black, line width=1pt] (0,0) ++(240:1.2cm) arc [start angle=240, end angle=300, radius=1.2cm];
     \draw[black, line width=1pt] (0,0) ++(-5:1.2cm) arc [start angle=-5, end angle=55, radius=1.2cm];

    \draw[blue, dashed, line width=1pt] (0,0) ++(55:1.2cm) arc [start angle=55, end angle=125, radius=1.2cm];
    \draw[blue, dashed, line width=1pt] (0,0) ++(300:1.2cm) arc [start angle=300, end angle=355, radius=1.2cm];
    \draw[blue, dashed, line width=1pt] (0,0) ++(185:1.2cm) arc [start angle=185, end angle=240, radius=1.2cm];

    \draw[blue, line width=1pt] ({1.2*cos(125)}, {1.2*sin(125)} ).. controls (-0.55,0.32)..(-0.6,0.3);
    \draw[blue, line width=1pt] (-0.65,0.2).. controls (-0.9,0.03)..({1.2*cos(185)}, {1.2*sin(185)});

    \draw[red, line width=1pt] ({1.2*cos(158)}, {1.2*sin(158)} )..controls (-0.9,0.45)..(-0.6,0.3);
    \draw[red, line width=1pt] ({1.2*cos(164)}, {1.2*sin(164)} )..controls (-0.9,0.35)..(-0.65,0.2);

    \draw[blue, dashed, line width=1pt] ({1.2*cos(121)}, {1.2*sin(121)} ).. controls (-0.4,0.2)..({1.2*cos(189)}, {1.2*sin(189)});
    \draw[blue, line width=1pt] ({1.2*cos(240)}, {1.2*sin(240)} ).. controls (0,-0.6)..({1.2*cos(300)}, {1.2*sin(300)});
    \draw[blue, dashed, line width=1pt] ({1.2*cos(236)}, {1.2*sin(236)} ).. controls (0,-0.5)..({1.2*cos(304)}, {1.2*sin(304)});
    \draw[blue, line width=1pt] ({1.2*cos(355)}, {1.2*sin(355)} ).. controls (0.5,0.2)..({1.2*cos(55)}, {1.2*sin(55)});
    \draw[blue, dashed , line width=1pt] ({1.2*cos(351)}, {1.2*sin(351)} ).. controls (0.4,0.2)..({1.2*cos(59)}, {1.2*sin(59)});

    \fill[green]
    (4,0) ++(-5:1.2cm)
    arc[start angle=-5, end angle=55, radius=1.2cm]

    .. controls (3.8,0) ..
     ({4+1.2*cos(240)}, {1.2*sin(240)})

     arc[start angle=240, end angle=300, radius=1.2cm]

    .. controls (4.6,-0.4) ..
    ({4+1.2*cos(355)}, {1.2*sin(355)})

    -- cycle;
    
    \draw[black, line width=1pt] (4,0) ++(125:1.2cm) arc [start angle=125, end angle=158, radius=1.2cm];
     \draw[black, line width=1pt] (4,0) ++(164:1.2cm) arc [start angle=164, end angle=185, radius=1.2cm];
     \draw[black, line width=1pt] (4,0) ++(240:1.2cm) arc [start angle=240, end angle=300, radius=1.2cm];
     \draw[black, line width=1pt] (4,0) ++(-5:1.2cm) arc [start angle=-5, end angle=55, radius=1.2cm];

    \draw[blue, dashed, line width=1pt] (4,0) ++(55:1.2cm) arc [start angle=55, end angle=125, radius=1.2cm];
    \draw[blue, dashed, line width=1pt] (4,0) ++(300:1.2cm) arc [start angle=300, end angle=355, radius=1.2cm];
    \draw[blue, dashed, line width=1pt] (4,0) ++(185:1.2cm) arc [start angle=185, end angle=240, radius=1.2cm];

    \draw[blue, line width=1pt] ({4+1.2*cos(125)}, {1.2*sin(125)} ).. controls (4-0.55,0.32)..(4-0.6,0.3);
    \draw[blue, line width=1pt] (4-0.65,0.2).. controls (4-0.9,0.03)..({4+1.2*cos(185)}, {1.2*sin(185)});

    \draw[red, line width=1pt] ({4+1.2*cos(158)}, {1.2*sin(158)} )..controls (4-0.9,0.45)..(4-0.6,0.3);
    \draw[red, line width=1pt] ({4+1.2*cos(164)}, {1.2*sin(164)} )..controls (4-0.9,0.35)..(4-0.65,0.2);

    \draw[blue, dashed, line width=1pt] ({4+1.2*cos(121)}, {1.2*sin(121)} ).. controls (4-0.4,0.2)..({4+1.2*cos(189)}, {1.2*sin(189)});

    \draw[blue, line width=1pt] ({4+1.2*cos(300)}, {1.2*sin(300)} ).. controls (4.6,-0.4)..({4+1.2*cos(355)}, {1.2*sin(355)});
    \draw[blue, dashed, line width=1pt] ({4+1.2*cos(304)}, {1.2*sin(304)} ).. controls (4.7,-0.4)..({4+1.2*cos(351)}, {1.2*sin(351)});
    \draw[blue, line width=1pt] ({4+1.2*cos(240)}, {1.2*sin(240)} ).. controls (3.8,0)..({4+1.2*cos(55)}, {1.2*sin(55)});
    \draw[blue, dashed, line width=1pt] ({4+1.2*cos(236)}, {1.2*sin(236)} ).. controls (3.8,0.1)..({4+1.2*cos(59)}, {1.2*sin(59)});

\end{tikzpicture}
\caption{\footnotesize Topologically distinct contributions to the kernel ${\cal I}(3)$. The two connected geometries in ${\cal I}_3(3)$, sliced through by the geodesic (in red), are topologically the same but differ in the number of boundaries in each slice. Nevertheless they contribute identically to the result.}
\label{I3}
\end{figure}
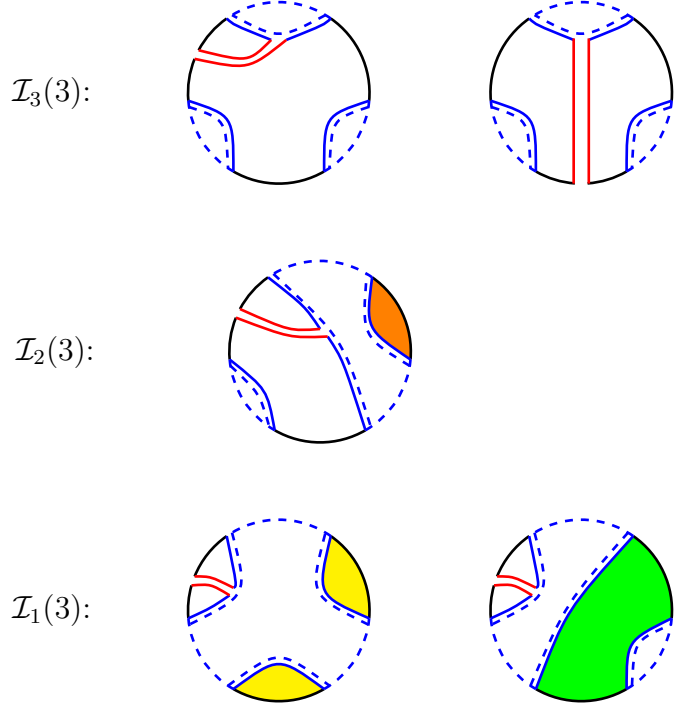
This case  is instructive as it illustrates two types of geodesics that can arise in the fully connected sector  ${\cal I}_3(3)$, given by  \eqref{n=3, L,1}, displayed at the top of Figure \ref{I3}. One slicing divides the geometry into a single AdS boundary half-disc and a  3-boundary  half-disc, whilst the second slicing splits it into two 2-boundary states. In the large time regime when the two-point spectral correlator $\braket{\rho(E_1)\rho(E_2)}$ is dominated by the coincidence limit $E_1\to E_2$, these different configurations yield identical contributions. This is verified explicitly in appendix \ref{AppendixE1}.

${\cal I}_2(3)$ given by eq. \eqref{n=3, L,2} accounts for all geodesics built in the second-most connected geometry, consisting of the $j=2$ replica wormhole, and the un-sliced disconnected disc $Z_1$, shaded in orange in Figure \ref{I3}.  Finally we have  ${\cal I}_1(3)$ accounting for all  geodesics that can be built cutting through a single disc. In this case the remaining two boundary-brane pairs are  either disconnected so we have two discs ($Z_{1}^{2}$ factor shaded  yellow in Figure \ref{I3}), or the replica wormhole connected geometry $Z_{2}$ (depicted in green in Figure \ref{I3}). The latter has one additional  EoW brane flavor loop, explaining the additional factor of $k$ accompanying $Z_2$ in \eqref{n=3, L,3}.
\subsection{General $n$ result}
For general $n$, it is easy to uncover the pattern in the expansion of ${\cal I}(n)$. In particular, the combinatorics associated to all ways of partitioning the unsliced portion of the geometry with $m$ boundaries is captured by the $m$-th complete exponential Bell polynomial \cite{Bell, Comtet},\footnote{The first few Bell polynomials are $B_0=1$, $B_1(x_1)=x_1$, $B_2(x_1,x_2) =(x_1^2 + x_2)$, $B_3(x_1,x_2,x_3)=x_1^3 + 3 x_1x_2+x_3$.}
\be
 B_{m}(x_{1},\dots,x_{m})=m!\sum_{\{j_p\geq 0\}}\;\prod_{p=1}^{m}\,\frac{x_{p}^{j_{p}}}{(p!)^{j_{p}}\,j_{p}!}\,,
\ee
with the $\{j_p\}$ constrained so they are associated to partitions of the integer $m$,
\be
\sum_{p\geq1} p\, j_p =m\,.
\ee
The complete Bell polynomial  represents the sum over all ways of  partitioning a set of size $m$ into non-empty, unordered subsets of  size $p$ and multiplicity $j_{p}$, each with weight $x_{p}$. In our context the weights $x_p$ must be identified with the connected $p$-boundary partition function $Z_p$ along with the index loop weighting factor,
\be
x_p = k^{p-1}\,Z_p\,.
\ee
Recalling from \eqref{Zn} that $y(E)\equiv e^{-\beta E}\gamma_{\mu}(E)$, the kernel ${\cal I}(n)$ assumes a compact form in terms of the Bell polynomials,
\begin{equation}\label{In}
\boxed{
    {\cal I}(n)=\sum_{p=1}^{n}y(E)^{p}\,k^{p}\,p^{2}\,\binom{n}{p}B_{n-p}(k^{0}Z_{1},k^{1}Z_{2},\dots,k^{n-p-1}Z_{n-p})\,.}
\end{equation}
However, we still need to understand how to perform the sum and obtain a result analytic in $n$ so that we can evaluate ${\cal I}^\prime(0)$, which is in turn required to calculate the expectation value of the length of the black hole interior, \footnote{Let us recall $s^{2}=2E$.} 
\begin{equation}\label{n-replica}
\overline{\braket{L_k(t)}}=L_{k,{\rm div}}-\frac{2\pi e^{2S_0}}{3k}\int_{s_{*}}^\infty ds\;
\frac{\partial{{\cal I}(n)}}{\partial n}\Bigg|_{n=0}
\rho^2_D(s)\, 
\left(1-\frac{st}{2 \pi  \rho_D(s)  e^{S_0}}\right)^3\,.
\end{equation}
We will address this below, first within the microcanonical ensemble, and then more non-trivially in the canonical picture, finding mutually consistent results.
\section{Microcanonical ensemble}\label{microcanonical}
Let us first look at the  computation of the geodesic length $\overline{\braket{L(t)}}$ in the microcanonical ensemble, meaning that we fix the energy in each asymptotic region, rather than fixing the renormalised boundary length $\beta$. Specifically, we consider states which have energies in a small band $\Delta E  \approx s\Delta s$ with $s=\sqrt{2 E}$ fixed.   It is then natural to define the microcanonical versions of the entropy and  $p$-boundary partition function which we denote using boldface symbols:
\begin{equation}
e^{\bf S}\, = \,e^{S_{0}}\rho_{D}(s) \Delta s, \hspace{20pt} {\bf Z}_p = e^{S_{0}}\rho_{D}(s)\Delta s\,y(s)^p\,,
\end{equation}
where $y(s)= e^{-\beta s^2/2} \gamma_\mu(s)$, so we have the microcanonical relation,
\be
{\bf Z}_p =e^{\bf S} \,y(s)^p\,.
\ee
\subsection{Kernel ${\cal I}(n)$ in microcanonical ensemble}
Then the complete Bell polynomials, appearing in ${\cal I}(n)$ \eqref{In} are
\begin{equation}
    B_{n-p}\left({\bf Z}_1,k^{1}{\bf Z}_2,\dots,k^{n-p-1}{\bf Z}_{n-p}\right)=B_{n-p}(a (ky(s)), \dots, a (ky(s))^{n-p})\,,
\end{equation}
where we have defined,
\begin{equation}\label{a/ratio}
    a\equiv \frac{e^{\bf S}}{k}\,,
\end{equation}
namely the ratio of the dimensions of the black hole and radiation Hilbert spaces. It is this ratio, more precisely its logarithm, that determines the onset of Page time for an evaporating black hole. The complete Bell polynomials are weighted isobaric polynomials,\footnote{An isobaric polynomial is one where each monomial has the same weight (sum of subscript indices).} which means that the following property holds,
\begin{equation}\label{property-Bell}
    B_{n-p}(a (ky), \dots, a (ky)^{n-p})\,=\,(ky)^{n-p}\, B_{n-p}(a,\dots, a)\,.
\end{equation}
 The complete Bell polynomial $B_p$ with all arguments identical is also known as the Touchard polynomial $T_p(a)$, which admits the following representation \cite{touchardref},
\begin{equation}
    T_p(a)=e^{-a}\sum_{r=0}^{+\infty}\frac{a^{r}\,r^{p}}{r!}\,.
\end{equation}
This displays an immediate relation to the Poisson distribution. $T_p(a)$ is the $p$-th moment of a Poisson distributed random variable with mean value $a$: $T_p(a) = {\mathbb E}[X^p]$, $X\sim$ Poisson(a).\footnote{Touchard polynomials have already appeared as moments of a partition-like random variable counting the dimension of asymptotically AdS Hilbert space in the topological toy model studied in \cite{Marolf:2020xie}.} With these ingredients in place, the kernel ${\cal I}(n)$ simplifies considerably,
\begin{equation}
    {\cal I}(n)
    =\,n\,(ky(s))^{n}\,e^{-a}\,\sum_{r=0}^{+\infty}\left[\frac{a^{r}}{r!}\,\frac{(n+r)}{\left(1+r\right)^{2-n}}\right].
\end{equation}
Viewing ${\cal I}(n)$ as an analytic function of $n$, ${\cal I}'(0)$ is then straightforward to  evaluate as an infinite sum\footnote{The sum in question is ${\cal I}'(0) = e^{-a}\sum_{r=0}^{+\infty}\left[\frac{a^{r}}{r!}\frac{r}{(1+r)^{2}}\right]$.} which can in turn be expressed in terms of the exponential integral function,
\begin{equation}\label{I'(0)-micro}
    \frac{\partial{{\cal I}(n)}}{\partial n}\Bigg|_{n=0}
    =\frac{e^{-a}}{a}\left(e^{a}+\gamma -\text{Ei}(a)+\log a-1\right)\,,
\end{equation}
where $\gamma$ is the Euler-Mascheroni constant. Equivalently, utilising the series representation of the right hand side, ${\cal I}'(0)$ can be viewed as the expectation value of the function $f(X)=X/(1+X)^2$ of a Poisson distributed random variable $X$ with mean value $a$, 
\begin{equation}\label{derivativeI}
{\cal I}'(0)
=\mathbb{E}\left[ \frac{X}{(1+X)^2} \right]\,,
\qquad {X\sim \text{Poisson}(a)}\,.
\end{equation}
\begin{figure}
    \centering
    \includegraphics[width=0.55\linewidth]{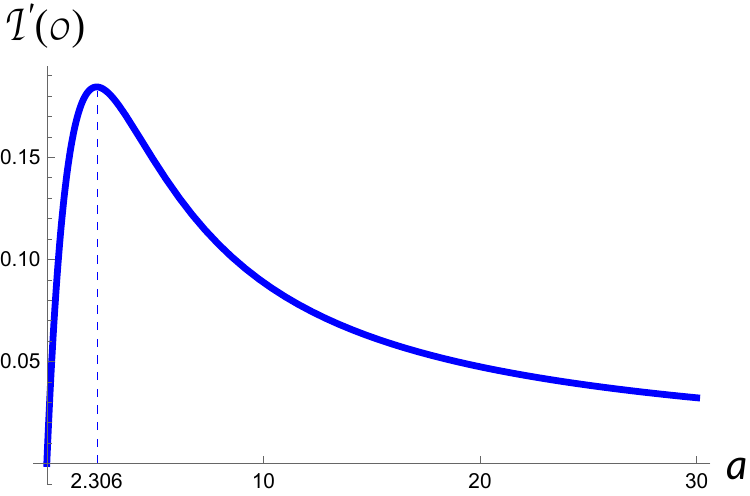}
    \caption{${\cal I}'(0)$ as a function of the ratio $a= e^{\bf S}/k$.}
\end{figure}
The most interesting feature of ${\cal I}'(0)$ as a function of the ratio $a$ is that it is a non-monotonic function with a maximum at $a_{\rm max} \approx 2.306$. This is when the radiation and black hole entropies are essentially equal i.e. ${\bf S} = \log k  + {\cal O}(1)$ and therefore should be identified with the onset of Page time up to small corrections. Whilst  $k$ is just a tunable parameter from the perspective of the EoW brane model, in an evaporating black hole, the semiclassical radiation entropy $S_{\rm rad}=\log k$ grows linearly with time. Therefore, $\log k$ should be viewed as a proxy for time in a slowly evaporating black hole. We will be more precise about this below.

The asymptotic expansions of ${\cal I}'(0)$ for small and large $a$ \footnote{Near $a=0$, ${\cal I}'(0)$ is regular and admits a power series expansion:\be
{\cal I}'(0)\big|_{a\ll1}= \frac{a}{4} - \frac{5 a^2}{36} +{\cal O}(a^3)\,.
\ee
The large $a$ behaviour on the other hand, apart from the leading $1/a$ term, is dictated in part by the asymptotic series for ${\rm Ei}(a)$,
\be
{\cal I}'(0)\big|_{a\gg 1} = \frac{1}{a}-\sum_{p=0}^\infty \frac{p !}{a^{p+2}} + e^{-a}(\log a -1 + \gamma)/a\,.
\ee
}
yield the  following pre-Page and post-Page time (``time" refers to $\log k$) behaviours,
\begin{equation}
    \frac{\partial{{\cal I}(n)}}{\partial n}\Bigg|_{n=0}\simeq\begin{cases}
    {\displaystyle \frac{k}{e^{\bf S}}+{\cal O}(k^2 e^{-2\bf S}) +  \mathcal{O}\!\left(e^{-\frac{e^{\bf S}}{k}}\,({\bf S}-\log{k})\,k{e^{-\bf S}}\right)} & \quad e^{\bf S} \gg k\,,\\[6pt]
    \\
    {\displaystyle \frac{e^{\bf S}}{4k} + \mathcal{O}\left(e^{2\bf S}{k^{-2}}\right)} & \quad  e^{\bf S} \ll k\,.
\end{cases}
\end{equation}
\subsection{The geodesic length}
Now let us understand what this means for the expectation value of the geodesic length in the microcanonical ensemble. In practice, this means restricting the energy integral \eqref{n-replica} to the narrow energy window of $\delta E \approx s\Delta s $ , and noting also that the integral has non-vanishing support only  when $s> s_*(t)$,  
\begin{equation}\label{expectation_L_2}
\overline{\braket{L_k(t)}}=L_{k, {\rm div}}-\frac{2\pi}{3}\Theta(s-s_{*}(t))\,\frac{\partial{{\cal I}(n)}}{\partial n}\Bigg|_{n=0} 
\frac{e^{\bf S}a}{\Delta s}\left(1-\frac{s\Delta s\,t}{2 \pi e^{ \bf S}}\right)^3\,.
\end{equation}
For $t\gg e^{S_{0}}$, it follows that  $s_{*}(t)\to +\infty$, \footnote{We recall $s_*(t)= \frac{1}{2\pi}\sinh^{-1}2\pi t$ which diverges logarithmically at late times, as reviewed in Appendix \ref{review-IMS}.} and the time dependent piece vanishes. In the canonical ensemble, the integrand vanishes exponentially as $s\to +\infty$, so this is smoothly implemented without any step functions. The length as defined has an additive ambiguity, and a natural renormalisation prescription which resolves these ambiguities is the one proposed in \cite{Iliesiu:2021ari},
\be
\overline{\braket{L_k(t)}}_{\rm ren}=\overline{\braket{L_k(t)}}-\overline{\braket{L_k(0)}}\,.\label{lren}
\ee

\subsubsection{Interior growth rate: fixed $k$}
One of the most striking consequences of \eqref{expectation_L_2} is the effect of  $k$ on the rate of growth of the interior at early times. This is given by the term linear in $t$ and is the ``velocity" of growth, 
\be
v_k(t)\equiv \left.\frac{d}{dt}\, \overline{\braket{L_k(t)}}_{\rm ren}\,\right|_{t\ll e^{S_0}}\,=\, {\cal I}'(0) \,a \,s\,=\begin{cases}s \qquad &e^{\bf S}\gg k\\\\
s\,{\displaystyle \frac{e^{2{\bf S}}}{4 k^2}} \qquad &e^{\bf S}\ll k\,.
\end{cases}
\,.
\ee
It is easily seen in the canonical ensemble at temperature $\beta^{-1}$ that $s\approx 2\pi/\beta$. Therefore, as the  dimension $k$ of the auxiliary Hilbert space is cranked up, the growth rate of the black hole interior decreases and becomes vanishingly small when the radiation Hilbert space is much larger than the black hole Hilbert space. This is a nontrivial consequence of entangling the black hole/EoW brane system with an  auxiliary radiation system.

\subsubsection{Evaporating scenario: $k\to k(t)$ }
Now let us turn to the implications of \eqref{expectation_L_2} for an evaporating black hole with $k$ a function of time. It is natural to assume that the system, for most times, evolves adiabatically so that a quasi-static or quasi-equilibrium picture applies.  The outgoing radiation can be approximated by a thermal spectrum at the instantaneous Hawking temperature $\beta^{-1}(t)$, while
the radiation Hilbert space $k(t)$  can be expressed in terms of the \emph{semiclassical} Hawking radiation
entropy $S_{\mathrm{rad}}(t)$ as
\begin{equation}
    k\to k(t) = e^{S_{\mathrm{rad}}(t)}\,.
\end{equation}
For instance, we can assume that the radiation bath is a $(1+1)$-d CFT with central charge $c$, and the entropy flux of radiation assumes the universal form (see e.g. \cite{Hollowood:2020cou})
\begin{equation}
    \dot{S}_{\text{rad}}(t)=\frac{\pi c}{6\beta(t)}\,,
\end{equation}
under the adiabatic condition $|\dot{\beta}(t)|\ll 1$, so one can then write,
\begin{equation}
    S_{\mathrm{rad}}(t)
    \simeq \dot S_{\mathrm{rad}}\,t\,.
\end{equation}
Given the above picture, we will assume that time evolution in a dynamical setting can be captured by a sequence of fixed-$k,\beta$ models, with the sequence of parameter values chosen to mimic the expected semiclassical time evolution. 

Then we naturally expect that the  renormalised geodesic length operator is obtained from \eqref{lren} by the replacement $k\to k(t)$,
\be
\overline{\braket{L_{k(t)}(t)}}_{\rm ren}=\overline{\braket{L_{k(t)}(t)}}-\overline{\braket{L_{k(t)}(0)}}\,.
\ee
In this instance, it is worth remarking that the prescription is not just a choice but in fact forced upon us. The reason is that  divergent contributions $L_{k,{\rm div}}$ in \eqref{expectation_L_2} depend nontrivially on $k$ and ${\bf S}$ which will both be time dependent in the sense explained  above, and we must therefore regulate the divergences at each time step of evaporation. The obvious choice would have been to subtract off the length at $t=0$, $L_{k(0)}(0)$, however this would fail to work at times $t>0$ when $k\neq k(0)$.

Since the interior volume of the black hole is expected to be a measure of the quantum complexity of the state, we propose that our renormalised length at time $t$ corresponds to the quantum complexity of the evaporating black hole subsystem,
\begin{equation}\label{renormalization}
    {\mathcal{C}}(t)=\overline{\braket{L_{k(t)}(t)}}_{\rm ren}\,.
\end{equation}
\begin{figure}[ht]
    \centering
    \includegraphics[width=0.69\linewidth]{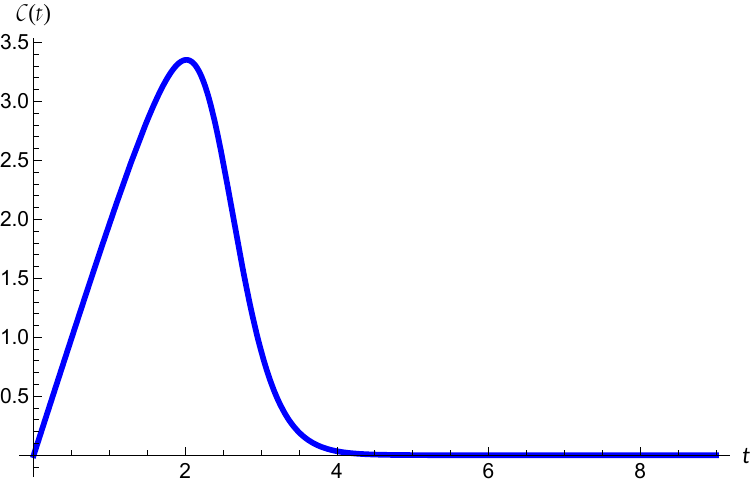}
    \caption{The renormalised geodesic length ${\cal C}(t)=\overline{\braket{L_{k(t)}(t)}}-\overline{\braket{L_{k(t)}(0)}}$ in the microcanonical ensemble, where  $S_{\text{BH}}$=6 and $S_{\text{rad}}(t)=\dot{S}_{\text{rad}}t$ with $\dot{S}_{\text{rad}}=2$. The geodesic length grows linearly, undergoes a transition (inflection point) at the Page time $t_{\text{Page}}=S_{\text{BH}}/\dot{S}_{\text{rad}}=3$ and decays exponentially.}
\end{figure}
For the moment we  self-consistently focus attention on the early time regime i.e. $t\ll e^{S_{0}}$, so we write,
\begin{equation}\label{micro_early3}
\mathcal{C}(t)\,=\,s\,\frac{e^{\bf S}}{k(t)}\frac{\partial{{\cal I}(n)}}{\partial n}\Bigg|_{n=0} 
t\;,\qquad \qquad t\ll e^{S_0}\,.
\end{equation} 
Expanding ${\cal I}'(0)$ as a function of $a = e^{\bf S}/k$ for $a\gg 1$, keeping the leading and first sub-leading corrections we find that ${\cal C}(t)$ starts growing linearly at first, but then reaches a maximum at $t=t_{\rm max}$,
\begin{equation}\label{max_micro}
 \dot{\cal C}(t_{\rm max})=0\,,  \qquad    t_{\rm max}\simeq\frac{{\bf S}-\log{{\bf S}}}{\dot{S}_{\text{rad}}}\,.
\end{equation}
This is a striking result. Beyond this point, the complexity undergoes a sharp transition to  exponential decay around, 
\begin{equation}
    t\simeq\frac{{\bf S}+\mathcal{O}(1)}{\dot{S}_{\text{rad}}}\,,
\end{equation}
which is an inflection point and can be identified with the Page time. The complete picture of the time evolution of ${\cal C}(t)$ can be summarized as below:
\begin{equation}
    \mathcal{C}(t)\simeq \begin{cases}
        st &\quad\quad\quad t\ll t_{\text{Page}}\\
        \frac{st}{4}\, e^{2(\mathbf{S}-\dot{S}_{\mathrm{rad}}\, t)}&\quad\quad\quad t_{\text{Page}}\ll t\lesssim e^{S_{0}} \\
        \frac{\pi}{6\Delta s}e^{2(\mathbf{S}-\dot{S}_{\mathrm{rad}}\, t)+{\bf S}}&\quad\quad\quad t\gg e^{S_{0}}\,.
    \end{cases}
\end{equation}
It is worth emphasizing that what we are computing here is a candidate  \textit{subsystem} complexity, via  the non-perturbative evaluation of the interior length. The results above are consistent with the recent findings on subsystem complexity \cite{Fan:2025moc, Haah:2025hyf}. Before  Page time, since the black hole is the larger of the two subsystems its complexity grows linearly, which is consistent with the expectation for early time growth of the complexity of the larger subsystem in a bipartite system. We have also found that the rate of this linear growth is  bounded from above by that of the non-evaporating black hole \cite{Iliesiu:2021ari} (the latter reviewed in Appendix \ref{review-IMS}). The subsystem complexity of the evaporating black hole reaches a maximum just before the Page time, when the Hilbert space dimensions of the black hole and radiation become comparable, after which we see a smooth  exponentially decrease. The drop is due to the fact that the black hole state becomes closer and closer to a maximally mixed state, which is  simple and contains almost no information, since it has no preferred basis and it is completely specified by the Hilbert space dimension.  The smooth post-Page time decrease is one of  two possible conjectured scenarios \cite{Haah:2025hyf}, the other being an abrupt fall to zero, and is potentially a signature of the fact that the entanglement wedge of the black hole does not change abruptly.
\section{Canonical ensemble}
\label{canonical}
We would like to now corroborate our observations based on the microcanonical treatment with a complete analysis of the geodesic integral \eqref{n-replica} in the canonical ensemble. This will be a two-step process -- first, obtaining an integral representation for the kernel ${\cal I}'(0)$ and then employing a saddle-point method to evaluate the renormalised length/complexity. 
\subsection{Integral representation of ${\cal I}'(0)$}
In the canonical ensemble, we need to work with the multi-boundary partition functions \eqref{Zn} which we now write as
\be\label{zhat}
Z_p= e^{S_0} \hat Z_p\,,\qquad \hat Z_p \equiv \int_0^\infty 
ds\,\rho_{D}(s) y(s)^{p}\,,
\ee
and introduce the parameter
\be\label{ao}
a_0\equiv\frac{e^{S_{0}}}{k}\,,
\ee
With these rescalings we can exploit the isobaric property \eqref{property-Bell} of Bell polynomials to write the kernel ${\cal I}(n)$ as \eqref{isoI'},
and to show (appendix \ref{AppendixE.2})  that ${\cal I}(n)$ is the coefficient of $z^n$ in the series expansion of an analytic function of $z$,
\be
{\cal I}(n)\, =\, k^{n}n![z^n]\, e^{y(s) z} \,y(s)\,z (1+y(s)z)\,B(z)\,,
\ee
where $B(z)$ is the generating function for  the complete Bell polynomials,
\begin{equation}\label{generating_Bell}
    B(z)=\exp{\left(a_0\sum_{p=1}^{\infty}\hat{Z}_{p}\,\frac{z^{p}}{
    p!}\right)}=\sum_{p=0}^{\infty} B_{p}(a_0\hat{Z}_{1},\dots,a_0\hat{Z}_{p})\frac{z^{p}}{p!}\,.
\end{equation}
After some work, we arrive at an integral representation for ${{\cal I}'(0)}$ which will play a key role in our saddle point evaluation,
\bea\label{integralrep}
\left.\frac{\partial {\cal I}(n)}{\partial n}\right|_{n=0}&&=y(s)\int_{0}^{\infty}d\chi\,\left(1-\chi y(s)\right) e^{-\chi y(s)} e^{-a_0\Phi(\chi)}\nonumber\\\\\nonumber
\Phi(\chi)&&=\int_{0}^{\infty} ds\,\rho_D(s)\left(1-e^{-\chi y(s)}\right)\,.
\eea
Recall from \eqref{Zn} that $y(s)$  is the Boltzmann factor times $\gamma_\mu(s)$, the modification in the density of states introduced by the EoW branes \eqref{EoWfunction}.

\subsection{Complexity evolution}
Let us  return to the expression for the  unrenormalised geodesic length,
\begin{equation}
    \overline{\braket{L_k(t)}}=L_{k,\rm div}-\frac{2\pi e^{S_0}a_{0}}{3}\int_{s_{*}}^\infty ds\;
{\cal I}'(0)
\;\rho^2_D(s)\, 
\left(1-\frac{s\,t}{2 \pi  \rho_D(s)  e^{S_0}}\right)^3\,.
\end{equation}
Our microcanonical analysis has already revealed that the early-time regime $t\ll e^{S_{0}}$ is the really interesting epoch for the  evolution of the interior length. Retaining only the linear  term in $t$ in the integrand, our prescription \eqref{renormalization}  for the renormalised length or complexity gives,
\begin{equation}\label{early_time_comp_can}
    \mathcal{C}(t)=t\,a_{0}\int_{0}^\infty ds\;s\,\rho_D(s)\,{\cal I}'(0)\,,\qquad t\ll e^{S_0}\,.
\end{equation}
To stay in an analytically tractable regime, we choose to work in the semiclassical or high temperature limit with $\beta \ll1$,
and further take the brane tension to be large $\mu\gg1/\beta$. Then, following \eqref{largemuy},
\begin{equation}
    y(s)\approx y(0)\,\hat y(s)\,,\qquad\quad y(0)\equiv\mu^{2\mu}\,,\qquad\quad \hat y(s)\equiv e^{-\frac{\beta s^{2}}{2}}\,.
\end{equation}
Since the integrand in our integral representation \eqref{integralrep} for ${\cal I}'(0)$  only depends on the combination $\chi y(s)$,  it is natural to absorb the factor of $y(0)$  into $\chi$ via the integration variable change $u=y(0)\chi$:
\begin{equation}\label{I0,can} 
    {\cal I}'(0)=\hat{y}(s)\int_{0}^{\infty}du\,\left(1-u \hat{y}(s)\right) e^{-u \hat{y}(s)}e^{-a_0\Phi(u)}\,,\qquad  u\equiv \chi\,y(0)\,.
\end{equation}
Likewise,
\begin{equation}\label{Phi(u)}
    \Phi(u)=\int_{0}^{\infty} ds'\,\rho_D(s')\left(1-e^{-u \hat{y}(s')}\right)\,.
\end{equation}
We note that $\Phi(u)$ is a non-negative, monotonically increasing function of $u$. In addition, $\Phi(0)=0$, hence for small $u$ it is linear,
\be
\Phi(u)\approx \Phi'(0) u\,,\qquad \Phi'(0)=\hat{Z}(\beta)=\frac{e^{\frac{2\pi^2}{\beta}}}{\sqrt{2\pi}\,\beta^{3/2}} \,.
\ee
In the linearised limit, the integrand is Gaussian and dominated by the saddle at $s_0=2\pi/\beta$, so we may na\"ively expect the linear behaviour to hold for $u\hat y(s_0)\ll1$, or $u\ll e^{2\pi^2/\beta}$.\footnote{The series expansion for $\Phi(u)= \frac{1}{\sqrt{2\pi}\beta^{3/2}}\sum_{n=1}^\infty(-1)^{n+1} \frac{u^n}{n!n^{3/2}} e^{2\pi^2/n\beta}$ is absolutely convergent. } 
Note that $\Phi'(0)$ is the disc partition function of JT gravity evaluated at the saddle. In the high temperature limit 
\be
\Phi'(0) \sim e^{2\pi^2/\beta}\gg1\,.
\ee
This has an important consequence for ${\cal I}'(0)$ in \eqref{I0,can}. In the linear regime for $\Phi(u)$, the integrand has a sharp maximum at,
\be
u_{\rm max}=\frac{1}{\hat y(s) + a_0\,\Phi'(0)}\,.
\ee
Given that $\hat y(s) \leq 1$, it is guaranteed for $a_0 \sim  {\cal O}(1)$, that $u_{\rm max}\sim {\cal O}(e^{-2\pi^2/\beta})$ and therefore the linear approximation for $\Phi(u)$ will work exceedingly well in the high temperature limit. However,  as we approach Page time $a_0\sim {\cal O}(e^{-4\pi^2/\beta})$,  $u_{\rm  max} \sim {\cal O}(e^{2\pi^2/\beta})$ which is close to the edge of validity of the linear approximation for $\Phi(u)$. At this point we must consider the effects outside the linear regime, as in Appendix \ref{canonical-details} and we find that they remain subleading at all times. Therefore, we find 
\be
{\cal I}'(0)\approx \frac{a_0\,\hat y(s) \Phi'(0)}{\left(\hat y(s) +a_0 \Phi'(0)\right)^2}\,.
\ee
The denominator  is a very slowly varying function of $s$ since $\hat y(s)$ is bounded above by unity, whilst the numerator is a Gaussian. Therefore ${\cal C}(t)$ can be evaluated readily by performing the Gaussian integral and evaluating the denominator at the saddle point,
\be
{\setlength{\fboxsep}{10pt}\boxed{\mathcal{C}(t)
    =\,\frac{2\pi}{\beta}\,t\,\,\frac{e^{2S_{\rm BH}(\beta)}}{\left(k(t)\,+\,e^{S_{\rm BH}(\beta)}\right)^{2}}}.}\label{comp_canonical}
\ee
Here we used the fact that the log-corrected Bekenstein-Hawking entropy of a black hole in JT gravity \cite{Maldacena:2016upp} is,\footnote{Let us recall that we have set $\phi_{b}=1$ and $8\pi G_{N}=1$ from the outset.}
\begin{equation}
    S_{\rm BH}(\beta)\simeq S_{0}+\frac{4\pi^{2}}{\beta}-\frac{3}{2}\log\beta\,\,,
\end{equation}
and we have restored $a_0=e^{S_{0}}/k$ and in the dynamical situation the  explicit time dependence of the radiation Hilbert space dimension $k(t)=e^{S_{\text{rad}}(t)}$. Indeed, in order to apply this to a slowly  evaporating near extremal black hole we must also replace $\beta$ with $\beta(t)$ in accordance with black hole thermodynamics and energy conservation. Given our present setup, we will reserve further detailed study of the full dynamical setting for future work.

Some comments on \eqref{comp_canonical} are in order. Remarkably, \eqref{comp_canonical} is reproduced, exactly in the pre-Page time phase and up a constant in the post-Page time phase,  by our microcanonical result \eqref{micro_early3} by replacing the Poisson random variable
$X$ by its mean
\begin{equation}
\mathbb{E}_{a(t)}\left[\frac{X}{(1+X)^2}\right] \rightarrow \frac{a(t)}{(1+a(t))^2}\qquad  \implies\qquad  \mathcal{C}(t)\big|_{\rm micro} \rightarrow \mathcal{C}(t)\big|_{\rm can}
\end{equation}
where $a(t)=e^{S_{\rm BH}}/k(t)$. This replacement is justified at early,
pre-Page times, where $a(t)\gg 1$. Indeed, the ratio of the standard deviation to the mean is $1/\sqrt{a}$ for a Poisson variable, thus the distribution of $X$ becomes sharply localised around its mean, and
the microcanonical average reduces to the canonical result. The post-Page time evolution, as we elaborate on further below, will see large fluctuations.

In the non-dynamical setting, dialling $k$ reduces the rate of complexity  growth or the ``velocity" from $2\pi/\beta$ for small $k$ to $\sim e^{2S_{\rm BH}}/k^2$ for large $k$ exactly as we found in the microcanonical ensemble.
The time evolution of \eqref{comp_canonical} is also consistent with the behaviour found in the microcanonical ensemble. In particular, we find a peak at time, 
\begin{equation}
    t_{\text{peak}}\simeq \frac{S_{\rm BH}(\beta)-\log S_{\rm BH}(\beta)}{\dot{S}_{\text{rad}}}\,,
\end{equation}
and a sharp transition (inflection point) at Page time,
\begin{equation}
    t_{\text{Page}}\simeq \frac{S_{\rm BH}(\beta)}{\dot{S}_{\text{rad}}}\,.
\end{equation}

\subsection{Variance of complexity}
The variance of complexity can be characterized via
\begin{equation}
    \sigma_{\bar{L}}^{2}(t)=\overline{\braket{L_k(t)^{2}}}-\overline{\braket{L_k(t)}}^{2}=\overline{\braket{L_k(t)^{2}}}_{\text{conn.}}
\end{equation}
where the subscript denotes that only connected two-point functions are taken into account in the evaluation of the expectation value of the geodesic length squared. Exploiting the identity
\begin{equation}
    L^{2}=\log^{2}{(e^{-L})}=\lim_{\Delta\to 0}\frac{\partial^{2}}{\partial\Delta^{2}}e^{-\Delta L}\,,
\end{equation}
it is straightforward to get the variance of complexity simply by following the same procedure based on the two-point function computation. Taking the second derivative of $\mathcal{M}(\Delta, E_{1},E_{2})$ in \eqref{M}  with respect to $\Delta$ yields,
\begin{equation}
\lim_{\Delta\rightarrow 0}\frac{\partial^{2}}{\partial\Delta^{2}} {\cal M}(\Delta,E_1,E_2)=\frac{ \sqrt{2 E} }{2\pi \gamma_{\mu}(E)
\rho_{D}(E)} \frac{F(E,\mu)}{\omega ^2}+\mathcal{O}(\omega^{-1})\,,
\end{equation}
in the $E_1\to E_2$ limit with $\omega=E_1-E_2$.
Thus, we can conclude that the $\omega$-integral in \eqref{eq:Lexpectation} remains unchanged and we get an additional function $F(E,\mu)$, containing hypergeometric and polygamma functions, in the final energy integral. For large brane tension $\mu\gg 1/\beta$ limit, the new factor $F(E,\mu)$ simplifies considerably,
\begin{equation}
    F(E,\mu)\simeq -2(\gamma+\log{4}) +\psi(2i\sqrt{2E})+\psi(-2i\sqrt{2E})\,.
\end{equation}
We find that the resulting noise-to-signal ratio, or the relative fluctuation of the interior length obeys,
\begin{equation}
    \frac{\sigma_{\bar L}(t)}{\overline{\braket{L_{k(t)}(t)}}_{\rm ren}}\simeq \sqrt{\frac{2\log{(\beta^{-1})}}{\overline{\braket{L_{k(t)}(t)}}_{\rm ren}}}\,.
\end{equation}
Consequently, for early times \(t \ll t_{\mathrm{Page}}\) the relative fluctuations are parametrically small, scaling as $ \beta^{1/2}\,\log^{1/2}\!\left(\beta^{-1}\right)\,\,t^{-1/2}$.
After the complexity peak, i.e.\,once the black hole interior starts to decay exponentially, the absolute fluctuations are exponentially suppressed as well, yet the relative fluctuations cease to be negligible. The latter become $\mathcal{O}(1)$ shortly after the Page-transition time and then exponentially increase.
One possible reason to explain this behaviour can be found by looking back at the microcanonical ensemble result: the complexity is controlled by an expectation value of a bounded nonlinear function of a Poisson-distributed variable $X$ with exponentially decaying mean \eqref{derivativeI}, and it turns out that the post-Page-time expectation value is dominated by exponentially rare configurations $X\simeq 1$. As a result, while the ensemble-averaged complexity decays exponentially, its relative fluctuations grow exponentially, signaling a loss of typicality.

\section{Conclusions}
\label{summary}
In this work we studied the evolution of the interior of an evaporating black hole in
JT gravity with an end-of-the-world brane. Evaporation was modelled by entangling
the internal states of the brane with an auxiliary radiation system, whose Hilbert
space grows during the evaporation process. We probed the black hole interior through
a boundary-to-brane two-point function and extracted from it a renormalised geodesic
length, which we interpreted as a candidate measure of the subsystem complexity of
the evaporating black hole.

The main new ingredient in the computation is the quenched ensemble average over
the brane data. This naturally leads to a replica prescription in which the geodesic
length receives contributions not only from the usual spacetime wormholes of pure JT
gravity, but also from replica wormholes associated with the radiation degrees of
freedom. These replica wormholes encode the correlations responsible for the Page
transition and modify the behaviour of the interior length in an essential way.

For a non-evaporating black hole, the renormalised geodesic length reproduces the
expected behaviour: it grows linearly at early times and approaches a plateau at
a time exponential in the entropy. In the evaporating case, however, we found a qualitatively
different evolution. The ensemble-averaged interior volume initially grows
linearly, reaches a maximum close to the Page time, and then decreases
exponentially. In this sense, the same non-perturbative gravitational effects that
restore the Page curve for the entropy also leave a sharp imprint on a geometric
observable probing the black hole interior.

This behaviour has a natural interpretation from the point of view of subsystem
complexity. Before the Page time, the black hole is the larger subsystem and its
complexity grows approximately as in the non-evaporating case, although with a rate
bounded by the corresponding eternal black hole result. Around the Page time, the
black hole and radiation Hilbert spaces become comparable in size. After this point,
the black hole subsystem becomes increasingly mixed, and the corresponding
subsystem complexity decreases rather than continuing to grow towards the usual
complexity plateau.

We also studied the fluctuations of the geodesic length. Before the Page time, the
relative fluctuations are parametrically small, so the ensemble average is
representative of a typical member of the ensemble. After the Page time, instead,
the mean complexity decreases exponentially while the relative fluctuations become
large. This signals a loss of self-averaging: the late-time ensemble average is no
longer controlled by typical configurations, but is dominated by rare members of the
ensemble. 

There are several natural directions for future work. 
One possibility is to extend to
the evaporating setup the analysis of geodesic states performed for the
non-evaporating black hole \cite{Iliesiu:2024cnh}. In particular, it would be interesting to study their
over-completeness and inner products, since these quantities should now contain
additional information associated with the EoW-brane flavour degrees of freedom and
with their evolution during evaporation. Another useful diagnostic would be the
wormhole velocity for individual members of the ensemble, rather than only after
ensemble averaging. This may provide a more refined probe of the interior geometry
and could help clarify the relation between complexity growth, rare configurations,
and possible firewall-like behaviour \cite{Iliesiu:2024cnh, Blommaert:2024ftn, Stanford:2022fdt, Franken:2026bff}. 

A key driving motivation for this work was to understand in what sense the  geometry inside a post-Page time black hole is, or is not, semiclassical. The interior length is clearly one diagnostic, and our analysis indicates that quantum effects on the interior geometry are unsuppressed post-Page time, paving the way for more direct probes of these effects through the experience of an infalling observer system using a microscopic dual framework such  as SYK coupled to a bath {\em and} an observer  along the lines of \cite{deBoer:2022zps, Gao:2021tzr, Bragagnolo:2025uia}.

It would also be valuable to connect the gravitational calculation presented here more directly with
quantum-information models of black hole evaporation. The appearance of combinatorial sums over weighted partitions which play a central role in our computation suggests a universal origin of the result. In view of the correspondence between
black holes and random quantum circuits \cite{Hayden:2007cs, Sekino:2008he}, one could ask whether a suitable random
circuit model, potentially selected among the proposals of \cite{Piroli:2020dlx, Gyongyosi:2023sue, Magan:2024aet}, reproduces the same pattern found here: early-time growth, a maximum
near the Page time, and a smooth post-Page decrease of subsystem complexity. Such
a comparison would help determine which aspects of our result are universal features
of evaporating chaotic systems and which depend on the specific JT gravity plus
EoW-brane ensemble considered in this work.

\acknowledgments 
The authors would like to thank Tim Hollowood, Carlos N\'u\~nez, Chanju Park and Andrea Legramandi for discussions. SPK would like to acknowledge support from STFC Consolidated Grant Award ST/X000648/1. NB is supported by the STFC DTP grant reference no. ST/Y509644/1. 
\\
\\
{\small {\bf Open Access Statement} - For the purpose of open access, the authors have applied a Creative Commons Attribution (CC BY) licence to any Author Accepted Manuscript version arising. 

Data access statement: no new data were generated for this work.}

\newpage

\begin{appendix}

\section{Genus expansion of pure JT gravity}
Pure JT gravity action \eqref{eq:EuclideanJTaction} can be rewritten as
\begin{equation}\label{path-integral-JT}
    I_{\text{JT}}=-S_0\chi(\mathcal{M}) - \int_{\mathcal{M}} \sqrt{g} \phi \left(R+2\right)
- 2 \int_{\partial \mathcal{M}} \sqrt{h} \phi \left(K-1\right)\,, \quad\quad 
\end{equation}
where the first term proportional to $S_{0}$ is purely topological and represents the Euler characteristic $\chi=2-2g-n$ of the Riemann surface $\mathcal{M}$, with $g$ the genus and $n$ the number of boundaries of $\mathcal{M}$. In particular, each time we add a handle to the Euclidean geometry, we increase the genus $g$ of the surface. These handles are often called spacetime wormholes\footnote{The reason for this nomenclature is to distinguish them from purely spatial wormholes that live at a fixed time-slice, such as the Einstein-Rosen bridge of an eternal black hole.} in this context.

The topological term allows us to represent the full path integral as a genus expansion\footnote{Multi-boundary correlators need to be computed by summing over all hyperbolic geometries ending on the union of all boundaries, and they can be either connected or disconnected. Whenever we put the subscript $C$, we are referring to the connected contributions.}
\begin{equation}\label{Ztimesn}
    \braket{Z(\beta_{1})\cdots Z(\beta_{n})}_{C}=\sum_{g=0}^{\infty}e^{(2-2g-n)S_{0}}Z_{g,n}(\beta_{1},\dots, \beta_{n})\,,
\end{equation}
where $Z_{g,n}$ denotes the contribution from geometries with fixed topology, specifically with $g$ handles.

Exploiting the fact that in a hyperbolic geometry we can always find geodesics homologous to each holographic (or Schwarzian) boundary, we can decompose each surface, associated to $Z_{g,n}$, into an inner hyperbolic surface $\mathcal{M}_{g,n}(\vec{b})$ of genus $g$ bounded by $n$ geodesics of lengths $b_{i}$ and $n$ trumpets, each describing a hyperbolic space bounded by a holographic boundary of renormalised length $\beta_{i}$ and a geodesic of length $b_{i}$.

Using this decomposition, it has been found \cite{Saad:2019lba}
\begin{equation}
    Z_{g,n}(\beta_{1},\dots,\beta_{n})=\int_{0}^{\infty}\left[\prod_{i=1}^{n}b_{i}db_{i}\,Z_{T}(\beta_{i},b_{i})\right]V_{g,n}(b_{1},\dots,b_{n})\,,
\end{equation}
where $V_{g,n}(b_{1},\dots,b_{n})=\text{Vol}(\mathcal{M}_{g,n}(\vec{b}))$ is the volume of the interior moduli space, which turns out to be induced by the Weil-Petersson symplectic form, and the integral is over the lengths of the geodesics homologous to each boundary. The remaining factors account for the partition function of the $n$ trumpet geometries, which is one-loop exact and given explicitly by
\begin{equation}
    Z_{T}(\beta,b)=\frac{e^{-\frac{b^{2}}{2\beta}}}{\sqrt{2\pi\beta}}\,.
\end{equation}
The additional $b_{i}$ factors are due to the integral over the moduli $\tau_{i}$, accounting for a possible twist occurring when the inner geodesic bounded geometry $\mathcal{M}_{g,n}(\vec{b})$ gets glued to the $n$ trumpets.

For the case $n=2$, the full genus expansion over connected contributions is not fully captured by \eqref{Ztimesn}, since the double trumpet, i.e. the topologically connected hyperbolic geometry with $n=2$ and $g=0$, must be treated separately, and it turns out that,
\begin{equation}
    Z_{0,2}(\beta_{1},\beta_{2})=\int_{0}^{\infty} bdb\,Z_{T}(\beta_{1},b)Z_{T}(\beta_{2},b)=\frac{1}{2\pi}\frac{\sqrt{\beta_1\beta_2}}{\beta_1+\beta_2}\;.
\end{equation}
Putting the ingredients together and performing an analytic continuation, one can derive the following key result for the spectral form factor,
\begin{equation}
    \braket{Z(\beta+it)Z(\beta-it)}=\int dE_1dE_2\,e^{-\beta(E_1+E_2)}e^{-it(E_1-E_2)}\braket{\rho(E_1)\rho(E_2)}\,.
\end{equation}
At late times, the integral is dominated by the coincident limit $E_{1}\to E_{2}$ of the density-density correlator 
$\braket{\rho(E_1)\rho(E_2)}$. At large $e^{S_0}$ and sufficiently away from the spectral edges, it has the following universal form for all matrix integrals with single-cut saddle points \cite{Saad:2019lba},
\bea \label{eq:sinekernel}
&&\braket{\rho(E_1) \rho(E_2)} = \\\nonumber
&&e^{2 S_0}\rho_D(E_1)
\rho_D(E_2)+e^{S_0}\rho_D(E_2)\delta(E_1-E_2)
-\frac{\sin^2\left(\pi e^{S_0}\rho_D(E_2)(E_1-E_2)\right)}{\pi^2 (E_1-E_2)^2}\,.
\eea

\section{Partition function of JT gravity with EoW brane}\label{AppendixA}
The fully connected disc partition function for JT gravity with $n$ AdS boundaries and $n$ EoW branes is
\begin{align}
Z_n = e^{S_0} \int ds\, \rho_D(s) y (s)^n\,,
\end{align}
with
\begin{equation}
    \rho_D(s)=\frac{s}{2\pi^{2}}\sinh{(2\pi s)}, \quad\quad\quad y(s)=e^{-{\beta s^2\over 2}}\left|\Gamma\left(\mu+\frac{1}{2}+is\right)\right|^2\,.
\end{equation}
The semiclassical limit is $\beta \ll 1$ corresponding to high temperatures. Upon restoring Newton's constant, it is equivalent   to the  $G_N \to 0$ limit.

For large brane tension $\mu\to\infty$, or more precisely $\mu\gg 1/\beta$ (the  partition function at high temperatures will only have support near $s\approx 2\pi/n\beta$), the product of Gamma functions is well approximated using  Stirling's formula as a $\mu$-dependent constant,
\begin{equation}\label{largemuy}
    y(s)\approx e^{-{\beta s^2\over 2}}\mu^{2\mu}\,.
\end{equation}
Therefore  to get the small-$\beta$ saddle point, we use 
\begin{align}
\rho_D(s)\, y(s)^n \approx\frac{s}{4 \pi^2}\, y(0)^n\, e^{2 \pi s - n\beta s^2/2}\,,
\end{align}
from which we get the saddle point value of $s$
\begin{align}
s^{(n)}_{\mu\gg1} = \frac{2 \pi}{n\beta} + {\cal O}(1).
\end{align}
Therefore, in the large brane tension limit, $Z_n$ is proportional to the thermal partition function of pure JT gravity at inverse temperature $n\beta$, and with a classical geometry given by the disc of boundary length $n\beta$.

In the tensionless brane limit, on the other hand, 
\begin{equation}
    y(s)=e^{-{\beta s^2\over 2}}{\pi\over\cosh \pi s}\approx2\pi  e^{-\pi s-{\beta s^2\over 2}}\,,
\end{equation}
and 
\begin{align}
\rho_D(s)\, y(s)^n \approx \frac{s}{ \pi}\, e^{(2-n)\pi s - n\beta s^2/2}\,,
\end{align}
with saddle point when $n<2$,
\begin{align}
s^{(n)}_{\mu\ll1} = \frac{(2-n) \pi}{n\beta} + {\cal O}(1).
\end{align}
Thus only  $Z_1$ has a classical (real) saddle which is a half disc. This means that the effective temperature is ${1\over 2\beta}$ and the classical geometry can be thought of as a $\mathbb{Z}_2$ quotient of the thermal disc.

\section{Review of complexity computation in JT with EoW brane}\label{review-IMS}
In this section we review the calculation of the non-perturbative geodesic length for JT gravity with an EoW brane (without attached degrees of freedom) carried out in \cite{Alishahiha:2022kzc} following on from the original work without EoW branes \cite{Iliesiu:2021ari}. Let us  start by writing the two-point function\footnote{In this section, by two-point function we mean  a correlation function between two operators, one  located at the asymptotic boundary and the other on the EoW brane.} at disc level,
\begin{align}
    &\left\langle\mathcal{O}_{1}(u)\mathcal{O}_{2}(0)\right\rangle_{D}=
    \frac{1}{Z_{D,\mu}(\beta)}\int_{0}^{\infty}\frac{dz}{z}\Psi_{D}(\beta-u,L)\Psi_{D}(u,L)\left(\frac{z}{4}\right)^{\Delta} \label{2p-disc-EoW}\\
    &=\frac{2e^{S_{0}}}{Z_{D,\mu}(\beta)}\int_{0}^{\infty}dE_{1}dE_{2}\;e^{-(\beta-u)E_{1}-uE_{2}}\gamma_{\mu}(E_{1})\gamma_{\mu}(E_{2})\rho_{D}(E_{1})\rho_{D}(E_{2})\mathcal{M}_{\Delta}(E_{1},E_{2})\,,\notag
\end{align}
where we have defined
\begin{equation}\label{M}
    \mathcal{M}_{\Delta}( E_{1},E_{2})=\int_{0}^{\infty}\frac{dz}{z^{2}}W_{-\mu,i\sqrt{2E_{1}}}(z)W_{-\mu,i\sqrt{2E_{2}}}(z)\left(\frac{z}{4}\right)^{\Delta}\,,\quad\quad\quad z=4e^{-L}\,.
\end{equation}
The disc-level two-point function \eqref{2p-disc-EoW} can be obtained by using the boundary particle formalism introduced in \cite{Yang:2018gdb}, according to which the full quantum gravity two-point function can be obtained by dressing the CFT two-point function $e^{-\Delta L}=(z/4)^{\Delta}$ with an appropriate kernel which, in this case, is the Hartle-Hawking wavefunction \eqref{HH-EoW}.

Higher genus and non-perturbative contributions can be included by performing the following substitution,
\begin{equation}\label{replacement}
    e^{2S_{0}}\rho_{D}(E_{1})\rho_{D}(E_{2})\to \braket{\rho(E_{1})\rho(E_{2})}\,,
\end{equation}
i.e. by replacing the leading disc contribution of the two-point correlation of the spectral density with its full non-perturbative expression \eqref{eq:sinekernel}. The extension, natural from a random matrix point of view, has been proven in \cite{Iliesiu:2021ari} to yield the full two-point function at any genus, evaluated on any non self-intersecting geodesics connecting the two insertions.

Now  we may analytically continue $u$ to real time, 
\begin{equation}
    u\to \frac{\beta}{2}+it\,,
\end{equation}
and extract the geodesic length from the full two-point function via,  
\bea
    \braket{L(t)}=-\lim_{\Delta\to0}\frac{\partial\braket{O\left(\frac{\beta}{2}+it\right)O(0)}}{\partial\Delta}\,.
\eea
All dependence of  the two-point function on $\Delta$ resides in  $\mathcal{M}_{\Delta}(E_{1},E_{2})$, so we can differentiate  with respect to $\Delta$ inside the energy integral. Then, using  ``center of mass" and relative variables,
\begin{equation}
    E=\frac{E_{1}+E_{2}}{2}, \quad\quad\quad \omega=E_{1}-E_{2}, \quad\quad\quad dE_{1}dE_{2}=dEd\omega\,,
\end{equation}
we find, using known integrals involving Whittaker functions \cite{gradshteyn2007},
\begin{equation}
\lim_{\Delta\rightarrow 0}\frac{\partial{\cal M}_{\Delta}(E_1,E_2)}{\partial\Delta}=\frac{ \sqrt{2 E} }{2\pi \gamma_{\mu}(E)
\rho_{D}(E)} \frac{1}{\omega ^2}+\mathcal{O}(\omega^{-1})\,.
\end{equation}
Combining these ingredients, we obtain
\begin{align}\label{eq:Lexpectation}
\braket{L(t)}=L_{\rm div}-&\frac{e^{S_0}}{\pi Z_{D,\mu}(\beta)}\int_0^\infty dE\,
e^{-\beta E}\,
\sqrt{2E}\;\gamma_\mu(E)\rho_{D}(E)\times\\ 
\times&\int_{-\infty}^\infty d\omega\,
\left(1
-\frac{\sin^2\left(\pi \rho_D(E) e^{S_0}\omega \right)}
{(\pi \rho_D(E)e^{S_0}\omega)^2}\right)e^{i\omega t}\left(\frac{1}{\omega^{2}}+\mathcal{O}(\omega^{-1})\right)\,.\label{finalfreq}
\end{align}
The contact term in \eqref{eq:sinekernel} leads to a divergent and $t$-independent contribution,  while the inclusion of subleading terms of order $\mathcal{O}(\omega^{-1})$ on the right hand side above, leads to similar  or vanishing contributions.

As emphasized in \cite{Iliesiu:2021ari}, the renormalization prescription for the length operator  is to consider the quantity $\braket{L(t)-L(0)}$, which is free of divergences and independent of regularisation scheme. Importantly, the renormalised length captures the expected late time physics of complexity evolution.

The final frequency  integral in \eqref{finalfreq}
can be evaluated exactly. The integrand is regular at $\omega=0$, due to the cancellation between the semi-classical disconnected piece and the non-perturbative contributions. However, we can evaluate it by expanding the sine in terms of complex exponentials and evaluating each contribution by contour integration, avoiding the putative pole in each term at $\omega=0$.  The $\omega$-integral is then performed  by closing the contour in the upper half-plane for $t> 2\pi  \rho_D(E)e^{S_0}$,  when the integral vanishes, since the contour does not enclose any poles. On the other hand, for $0< t< 2\pi \rho_D(E)e^{S_0}$,  we find a non-zero residue, 
\bea
&&\int_{-\infty}^\infty d\omega\,
\frac{e^{i\omega t}}{\omega^2}\,
\left(1
-\frac{\sin^2\left(\alpha_D\omega \right)}
{(\alpha_D \omega)^2}\right)\,,\qquad \alpha_D(E)\equiv\pi\rho_D(E)e^{S_0}\\\nonumber
&&=
\frac{2\pi\alpha_D}{3 }\, \left(1-\frac{t}{2 \alpha_D}\right)^3\,.
\eea
We finally obtain,
\begin{equation}
\braket{L(t)}=L_{\rm div}-\frac{2\pi e^{2S_0}}{3Z_{D,\mu}(\beta)}\int_{E_*}^\infty dE\,
e^{-\beta E}\,
\sqrt{2E}\;\gamma_\mu(E)\rho^2_D(E)\, 
\left(1-\frac{t}{2 \alpha_D(E)}\right)^3\,.
\end{equation} 
where $E_*$ is implicitly obtained via the equation $2\pi \rho_D(E_*)e^{S_0}=t$. Then the renormalised length or complexity exhibits linear growth, saturating at late times,
\begin{equation}
    \mathcal{C}(t)=\braket{L(t)-L(0)}\approx\begin{cases}
        c_{1} t+\dots &\quad\quad\quad t\ll e^{S_{0}}\,,\\
        c_{0}-\dots &\quad\quad\quad t\gg e^{S_{0}}\,,
    \end{cases}
\end{equation}
where $c_{0}\sim e^{S_{0}}$ and $c_{1}\simeq \frac{2\pi}{\beta}$ in the semiclassical and large brane tension  limit. The linear growth followed by a plateau, at a value and time exponential in the entropy of the black hole, is consistent with Susskind's conjecture \cite{Susskind:2015toa, Brown:2016wib, Susskind:2020wwe, Balasubramanian:2019wgd}.

\section{Page curve}\label{SR}
The fine-grained entanglement entropy of Hawking radiation can be extracted using the replica trick via
\begin{equation}
    S_{R}=\lim_{n\to1}\frac{1}{1-n}\log(\text{Tr}(\rho_{R}^{n}))=-\lim_{n\to1}\frac{\partial}{\partial n}\log\left(\text{Tr}(\rho_{R}^{n})\right)=-\lim_{n\to1}\frac{\partial}{\partial n}\text{Tr}(\rho_{R}^{n})\,,
\end{equation}
where 
\begin{equation}\label{Rényi}
    \text{Tr}(\rho_{R}^{n})=\frac{\overline{Z^{n}}}{(kZ_{1})^{n}}\,.
\end{equation}
In the planar limit, we can compute $\overline{Z^{n}}$ by considering all possible ways of filling in the bulk $n$ boundary conditions of the type shown in Figure \ref{bc-gampl}, neglecting both handles and brane crossing as explained in the main text.\\ 
It turns out that
\begin{equation}\label{Zn-EE1}
    \left(\overline{Z^{n}}\right)_{\text{planar}}= k\sum_{j=1}^{n}\sum_{\substack{r_1,\dots,r_n\geq0\\
                \sum_{m=1}^{n}mr_m=n\\
                \sum_{m=1}^{n}r_m=j}}\frac{n!}{(n-j+1)!\prod_{m=1}^{n}r_{m}!}\prod_{m=1}^{n}\left(k^{m-1}Z_{m}(\beta)\right)^{r_m}\,,
\end{equation}
where $r_m$ is the number of connected geometries with $m$ replica boundaries, yielding a total of $1\leq j \leq n$ connected blocks constructed out of $n$ pairs of asymptotic boundary/EoW brane. \\
The above complicated expression gets considerably simplified in the microcanonical ensemble, since we need to perform the following replacement
\begin{equation}
    Z_{m} \to {\bf Z}_{m}=e^{\bf S}y(E)^{m}\,,
\end{equation}
and \eqref{Zn-EE1} becomes
\begin{equation}\label{Zn-EE2}
    \left(\overline{{\bf Z}^{n}}\right)_{\text{planar}}=y(E)^{n}\sum_{j=1}^{n}k^{n-j+1}e^{{\bf S}j}\sum_{\substack{r_1,\dots,r_n\geq0\\
                \sum_{m=1}^{n}mr_m=n\\
                \sum_{m=1}^{n}r_m=j}}\frac{n!}{(n-j+1)!\prod_{m=1}^{n}r_{m}!}\,.
\end{equation}
Now the last sum in \eqref{Zn-EE2} gives the so-called Narayana numbers $N(n,j)$, counting the \textit{non-crossing} \footnote{It is worth remarking that in the replica trick applied to compute the ensemble averaged boundary-brane two-point function we did not consider non-crossing partitions, since there was not a global trace/cycle imposing it, as it is the case for the $n$-Rényi entropy \eqref{Rényi}.} partitions in which a set of $n$ elements is partitioned in exactly $j$ blocks, which get simplified into
\begin{equation}
    N(n,j)=\frac{1}{n}\binom{n}{j}\binom{n}{j-1}\,.
\end{equation}
Therefore the $n$-Rényi entropy can be written as
\begin{equation}
    \text{Tr}(\rho_{R}^{n})= \sum_{j=1}^{n}N(n,j)k^{j-n}e^{{\bf S}(1-j)}= \sum_{j=1}^{n}N(n,j)k^{1-j}e^{{\bf S}(j-n)}\,,
\end{equation}
specifically in two equivalent ways, related to the fact that \eqref{Zn-EE1} is symmetric under the $k\leftrightarrow e^{\bf S}$ swapping. However, as explained in \cite{Blommaert:2021etf}, the two equivalent expressions yield two analytic continuations to real $n$, each satisfying Carlson's theorem in different and complementary ranges, specifically
\begin{equation}
    \text{Tr}(\rho_{R}^{n})=k^{1-n}{}_2F_1\!\left(1-n, -n; 2; \frac{k}{e^{\bf S}}\right)\Theta(e^{\bf S}-k) +\;e^{{\bf S}(1-n)}{}_2F_1\!\left(1-n, -n; 2; \frac{e^{\bf S}}{k}\right)\Theta(k-e^{\bf S})\,.
\end{equation}
Finally we get the entanglement entropy, perfectly consistent with the leading large-dimension Page formula \cite{Page:1993df},
\begin{equation}
    S(R)= \left(\log{k}-\frac{k}{2 e^{\bf S}}\right)\Theta(e^{\bf S}-k)+ \left({\bf S}-\frac{e^{\bf S}}{2 k}\right)\Theta(k-e^{\bf S})\,.
\end{equation}
This derivation of the Page curve has been similarly found in \cite{Kudler-Flam:2021alo, Okuyama:2021ylc}.

\section{More on the complexity computation}
\subsection{Truncation of $\exp{(\#\omega\beta)}$ factors in the kernel ${\cal I}(n)$}\label{AppendixE1}
\begin{figure}[H]
\centering
\begin{tikzpicture}[baseline]
    \draw[black, line width=1pt] (0,0) ++(120:1.2cm) arc [start angle=120, end angle=190, radius=1.2cm] node[midway, left] {$\beta-u$};

    \draw[black, line width=1pt] (0,0) ++(190:1.2cm) arc [start angle=190, end angle=240, radius=1.2cm] node[midway, left] {$u$};

    \draw[black, line width=1pt] (2,0) ++(-60:1.2cm) arc [start angle=-60, end angle=60, radius=1.2cm] node[midway, right] {$\beta$};

    \draw[blue, line width=1pt] ({1.2*cos(120)}, {1.2*sin(120)} ).. controls (1,0.6)..({2+1.2*cos(60)}, {1.2*sin(60)});


    \draw[red, line width=1pt] ({1.2*cos(195)}, {1.2*sin(195)}) .. controls (0.6,-0.4).. (1,-0.7);

    \draw[orange, line width=1pt] ({1.2*cos(195)}, {1.2*sin(195)}) .. controls (0.2,-0.2).. (1,0.7);

    \draw[blue, dashed, line width=1pt] ({1.2*cos(120)}, {0.05+1.2*sin(120)} ).. controls (1,0.7)..({2+1.2*cos(60)}, {0.05+1.2*sin(60)});

    \draw[blue, dashed, line width=1pt] ({1.2*cos(120)}, {0.05+1.2*sin(120)} ).. controls (1,1.5)..({2+1.2*cos(60)}, {0.05+1.2*sin(60)});

    \draw[blue, dashed, line width=1pt] ({1.2*cos(120)}, {-0.05-1.2*sin(120)} ).. controls (1,-0.7)..({2+1.2*cos(60)}, {-0.05-1.2*sin(60)});

    \draw[blue, dashed, line width=1pt] ({1.2*cos(120)}, {-0.05-1.2*sin(120)} ).. controls (1,-1.5)..({2+1.2*cos(60)}, {-0.05-1.2*sin(60)});



    \draw[blue, line width=1pt] ({1.2*cos(120)}, {-1.2*sin(120)} ).. controls (1,-0.6)..({2+1.2*cos(60)}, {-1.2*sin(60)});



    \draw[black, line width=1pt] (8,0) ++(125:1.2cm) arc [start angle=125, end angle=158, radius=1.2cm] node[midway, left] {\footnotesize $\beta-u$};
     \draw[black, line width=1pt] (8,0) ++(158:1.2cm) arc [start angle=158, end angle=185, radius=1.2cm] node[midway, left] {\footnotesize $u$};
     \draw[black, line width=1pt] (8,0) ++(240:1.2cm) arc [start angle=240, end angle=300, radius=1.2cm];
     \draw[black, line width=1pt] (8,0) ++(-5:1.2cm) arc [start angle=-5, end angle=55, radius=1.2cm];

    \draw[blue, dashed, line width=1pt] (8,0) ++(55:1.2cm) arc [start angle=55, end angle=125, radius=1.2cm];
    \draw[blue, dashed, line width=1pt] (8,0) ++(300:1.2cm) arc [start angle=300, end angle=355, radius=1.2cm];
    \draw[blue, dashed, line width=1pt] (8,0) ++(185:1.2cm) arc [start angle=185, end angle=240, radius=1.2cm];

    \draw[blue, line width=1pt] ({8+1.2*cos(55)}, {1.2*sin(55)} ).. controls (8,0.6)..({8+1.2*cos(125)}, {1.2*sin(125)} );

    \draw[blue, dashed, line width=1pt] ({8+1.2*cos(189)}, {1.2*sin(189)} ).. controls (8-0.7,-0.35)..({8+1.2*cos(236)}, {1.2*sin(236)} );
    \draw[blue, line width=1pt] ({8+1.2*cos(355)}, {1.2*sin(355)} ).. controls (8+0.6,-0.3)..({8+1.2*cos(300)}, {1.2*sin(300)} );
    \draw[blue, dashed, line width=1pt] ({8+1.2*cos(351)}, {1.2*sin(351)} ).. controls (8+0.7,-0.35)..({8+1.2*cos(304)}, {1.2*sin(304)} );
    \draw[blue, line width=1pt] ({8+1.2*cos(185)}, {1.2*sin(185)} ).. controls (8-0.6,-0.3)..({8+1.2*cos(240)}, {1.2*sin(240)} );
    \draw[blue, dashed, line width=1pt] ({8+1.2*cos(59)}, {1.2*sin(59)} ).. controls (8, 0.7)..({8+1.2*cos(121)}, {1.2*sin(121)} );

    \draw[green, line width=1pt] ({8+1.2*cos(158)}, {1.2*sin(158)} ) .. controls (8,0)..(8+0.65,-0.35);
    \draw[violet, line width=1pt] ({8+1.2*cos(158)}, {1.2*sin(158)} ).. controls (8-0.4,0.4)..(8, 0.7);
    \draw[brown, line width=1pt] ({8+1.2*cos(158)}, {1.2*sin(158)} ) .. controls (8-0.75,-0.05)..(8-0.7,-0.35);

\end{tikzpicture}
\caption{Depiction of all naively inequivalent geodesics stretching between an asymptotic boundary and an EoW brane in a fully connected $n=2$ (left) and $n=3$ (right) replica wormhole geometry. The geodesics' colours match those of the corresponding Boltzmann weights in eq.\,\eqref{omega22} and eq.\,\eqref{omega33}.}
\end{figure}
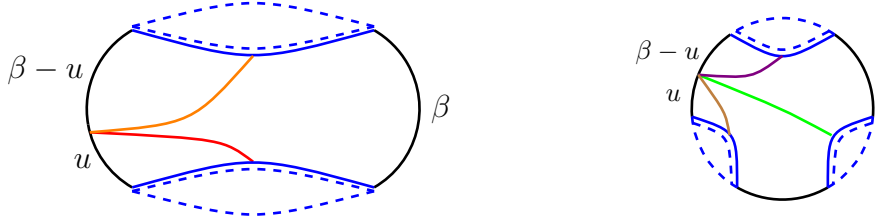
If we computed \eqref{OOn=2} keeping the $e^{\#\omega\beta}$ factors consistently, we would get 
\begin{align}
    &\big\langle\sum_{A=1}^{2}O_{1,A} \sum_{B=1}^{2}O_{2,B}\big\rangle\ \overline{Z^{2}} =\label{omega22}\\
&\qquad\qquad2e^{S_{0}}\int_{0}^{\infty}dEd\omega\Big\langle\rho\left(E+\frac{\omega}{2}\right)\rho\left(E-\frac{\omega}{2}\right)\Big\rangle\gamma_{\mu}(E)\mathcal{M}_{\Delta}(E,\omega)\times\nonumber\\\nonumber
    &\times\left[e^{-\beta E}\gamma_{\mu}(E)\,k\cdot 1\cdot1\cdot \binom{2}{1}Z_{1}\,+ e^{-2\beta E}\gamma_{\mu}(E)^{2}\,k^{2}\cdot 2\cdot\left({\color{orange}e^{\frac{\omega\beta}{2}}}+{\color{red}e^{-\frac{\omega\beta}{2}}}\right) \binom{2}{2}\right]\,,
\end{align}
and similarly, in the $n=3$ case,
\begin{align}
    &\big\langle\sum_{A=1}^{3}O_{1,A} \sum_{B=1}^{3}O_{2,B}\big\rangle\ \overline{Z^{3}} =
2e^{S_{0}}\int_{0}^{\infty}dEd\omega\Big\langle\rho\left(E+\frac{\omega}{2}\right)\rho\left(E-\frac{\omega}{2}\right)\Big\rangle\times\label{omega33}\\
&\qquad\qquad\qquad\gamma_{\mu}(E)\mathcal{M}_{\Delta}(E,\omega)\left[e^{-\beta E}\gamma_{\mu}(E)\,k\cdot 1\cdot1\cdot \binom{3}{1}(Z_{1}^{2}+kZ_{2})+\right.\nonumber\\
&\qquad\qquad\qquad\left.e^{-2\beta E}\gamma_{\mu}(E)^{2}\,k^{2}\cdot 2\left({\color{orange}e^{\frac{\omega\beta}{2}}}+{\color{red}e^{-\frac{\omega\beta}{2}}}\right)\binom{3}{2}Z_{1}\,+\right.\notag\\
&\qquad\qquad\qquad\left.
    +e^{-3\beta E}\gamma_{\mu}(E)^{3}\,k^{3}\cdot 3\left({\color{violet}e^{\omega\beta}}+{\color{green}e^{0}}+{\color{brown}e^{-\omega\beta}}\right)\binom{3}{3}\right]\,\nonumber.
\end{align}
Carrying out the prescription for general positive integer $n$, 
\bea
    {\cal I}(n)&&=\\\nonumber
    &&\sum_{j=1}^{n}(y(E))^{j}\,k^{j}j\sum_{m=0}^{j-1}\left[e^{\left(\frac{j-1}{2}-m\right)\omega\beta}\right]\binom{n}{j}B_{n-j}(k^{0}Z_{1},k^{1}Z_{2},\dots,k^{n-j-1}Z_{n-j})\,,
\eea
yielding, in the microcanonical ensemble, 
\begin{equation}
    \frac{\partial{{\cal I}(n)}}{\partial n}\Bigg|_{n=0}= e^{-a}\sum_{r=0}^{\infty}\left[\frac{a^{r}}{r!}\frac{e^{\frac{\omega\beta}{2}}r}{(e^{\frac{\omega\beta}{2}}+r)(1+e^{\frac{\omega\beta}{2}}r)}\right]= e^{-a}\sum_{r=0}^{\infty}\left[\frac{a^{r}}{r!}\frac{r}{(1+r)^{2}}\right]+\mathcal{O}(\omega^{2})\,.
\end{equation}
Similarly, in the canonical ensemble ${\cal I}'(0)$ \eqref{I'(0)-can} acquires $\mathcal{O}(\omega^{2})$ corrections.\\ 
Therefore, our truncation
\begin{equation}
    \sum_{m=0}^{j-1}\left[e^{\left(\frac{j-1}{2}-m\right)\omega\beta}\right]\simeq j\,,
\end{equation}
is well justified.

\subsection{Computation of ${\cal I}(n)$ in the canonical ensemble}\label{AppendixE.2}
Using the rescaled variables \eqref{zhat}, \eqref{ao} and the  isobaric property of the  complete Bell polynomials we re-express
\eqref{In} in a more malleable form:
\begin{equation}\label{isoI'}
    {\cal I}(n)=k^{n
}\sum_{p=1}^{n}y(s)^{p}p^{2}\binom{n}{p}B_{n-p}(a_0\hat{Z}_{1},a_0\hat{Z}_{2},\dots,a_0\hat{Z}_{n-p})\,.
\end{equation}
Given the generating function $B(z)$ for the complete Bell polynomials \eqref{generating_Bell}, let us also define the series
\begin{equation}
    A(z)\equiv\sum_{p=1}^{\infty}y(s)^{p}p^{2}\frac{z^{p}}{p!}=e^{y(s)z}\,y(s)z\,(1+y(s)z)\,.
\end{equation}
Exploiting the convolution property of two power series, according to which, given two power series
\begin{equation}
    A(z)=\sum_{m=0}^{\infty}{\cal A}_{m}\frac{z^{m}}{m!}, \quad\quad\quad B(z)=\sum_{m=0}^{\infty}{\cal B}_{m}\frac{z^{m}}{m!}\,,
\end{equation}
the product $C(z)=A(z)B(z)$ is a power series given by
\begin{equation}\label{convolution}
    C(z)= \sum_{n=0}^{+\infty}{\cal C}_{n}\frac{z^{n}}{n!}, \quad\quad\quad {\cal C}_{n}=\sum_{j=0}^{n}\binom{n}{j}{\cal A}_{j}{\cal B}_{n-j}\;,
\end{equation}
we can easily observe that ${\cal I}(n)$ \eqref{In} can be expressed as
\begin{equation}\label{In2_can}
    {\cal I}(n)=k^{n}n! [z^{n}]A(z)B(z)\;,
\end{equation}
where $[\cdots]$ denotes ``coefficient of $\cdots$".  $B(z)$ in \eqref{generating_Bell} has a compact form,
\begin{align}
    B(z)&=e^{-a_0M}\exp\left[a_0\int_{0}^{\infty} ds'\rho(s')e^{y(s')z}\right]\\
    &=e^{-a_0M}\sum_{r=0}^{\infty}\left[\frac{a_{0}^{r}}{r!}\int_{0}^{+\infty}\prod_{k=1}^{r}\Bigl(ds_{k}\rho(s_{k})\Bigr)e^{S_{r}z}\right]\,,
\end{align}
where we used \eqref{Zn} in the first equality, and defined
\begin{equation}
    M\equiv \int_{0}^{\infty} ds'\rho(s'), \quad\quad\quad\quad S_{r}\equiv \sum_{k=1}^{r}y(s_{k})\,.
\end{equation}
Then \eqref{In2_can} becomes
\begin{equation}
    {\cal I}(n)=e^{-a_0M}\sum_{r=0}^{\infty}\left[\frac{a_0^{r}}{r!}\int_{0}^{+\infty}\prod_{k=1}^{r}\Bigl(ds_{k}\rho(s_{k})\Bigr)\left(k^{n}n![z^{n}]A(z)e^{ S_{r}z}\right)\right]\,.
\end{equation}
Applying once again the series convolution property \eqref{convolution},
\begin{align}
    n![z^{n}]A(z)e^{ S_{r}z}&=\sum_{j=0}^{n}{\cal A}_{j}S_{r}^{n-j}\binom{n}{j}\\
    &=n S_{r}^{n-1}y(s)\left(1+\frac{y(s)}{S_{r}}\right)^{n-2}\left(1+n\frac{y(s)}{S_{r}}\right)\,,
\end{align}
we finally take the derivative with respect to $n$ and send $n\to 0$, obtaining the  intermediate result,
\begin{equation}\label{I'(0)-can}
    {\cal I}'(0)\equiv \frac{\partial I(n)}{\partial n}\Bigg|_{n=0}=e^{-a_0M}\,y(s)\sum_{r=0}^{+\infty}\left[\frac{a_0^{r}}{r!}\int_{0}^{+\infty}\prod_{k=1}^{r}\Bigl(ds_{k}\rho(s_{k})\Bigr)\frac{S_{r}}{(S_{r}+y(s))^{2}}\right]\,.
\end{equation}
At this point, we note that the last factor of the integrand can be written as a Laplace transform
\begin{equation}
    \frac{S_{r}}{(S_{r}+y(s))^{2}}=\int_{0}^{+\infty}d\chi\,\chi e^{-\chi(S_{r}+y(s))}S_{r}\,,
\end{equation}
so that \eqref{I'(0)-can} becomes
\begin{align}
    {\cal I}'(0)=&e^{-a_0M}\,y(s)\sum_{r=0}^{+\infty}\left[\frac{a_0^{r}}{r!}\int_{0}^{+\infty} d\chi\,\chi e^{-\chi y(s)}\times\right.\\
    &\left.\times\left(\sum_{j=1}^{r}\int_{0}^{+\infty} ds_{j}\,\rho(s_{j})y(s_{j})e^{-\chi y(s_{j})}\right)\left(\prod_{k\neq j}\int_{0}^{+\infty} ds_{k}\,\rho(s_{k})e^{-\chi y(s_{k})}\right)\right]\,,
\end{align}
or more compactly 
\begin{equation}\label{I'(0)-2}
    {\cal I}'(0)=e^{-a_0M}y(s)\int_{0}^{+\infty}d\chi\,\left(\chi e^{-\chi y(s)}\right)\left[\sum_{r=0}^{+\infty} r\frac{a_0^{r}}{r!}I_{1}(\chi)I_{0}(\chi)^{r-1}\right]\,,
\end{equation}
where we introduced the following notation
\begin{align}
    &I_{0}(\chi)=\int_{0}^{+\infty} ds\,\rho(s)e^{-\chi y(s)}\,,\\
    &I_{1}(\chi)=\int_{0}^{+\infty} ds\,\rho(s)y(s)e^{-\chi y(s)}\,.
\end{align}
Performing the $r$-series in \eqref{I'(0)-2}, we get
\begin{align}\label{I'(0)-4}
    {\cal I}'(0)&=e^{-a_0M}a_0\,y(s)\int_{0}^{+\infty}d\chi\,\left(\chi e^{-\chi y(s)}\right)I_{1}(\chi)e^{a_0I_{0}(\chi)}\\\notag
    &=y(s)\int_{0}^{+\infty}d\chi\,\left(1-\chi y(s)\right) e^{-\chi y(s)}\exp\left[a_0\int_{0}^{+\infty} ds''\,\rho(s'')\left(e^{-\chi y(s'')}-1\right)\right]\,,
\end{align}
where to get to the last step we performed an integration by parts recognising,
\begin{align}
    &-a_0\left(\int_{0}^{+\infty} ds'\,\rho(s')y(s')e^{-\chi y(s')}\right)\exp\left[a_0\int_{0}^{+\infty} ds''\,\rho(s'')\left(e^{-\chi y(s'')}-1\right)\right]\notag\\
    &=\partial_{\chi}\left(\exp\left[a_0\int_{0}^{+\infty} ds''\,\rho(s'')\left(e^{-\chi y(s'')}-1\right)\right]\right)\,.
\end{align}
Furthermore, let us note that if we go to the microcanonical ensemble, we get exactly \eqref{I'(0)-micro}, which represents a strong consistency check that the correct analytic continuation in $n$ has been carried out.
\subsection{Leading contribution of the canonical ensemble computation}\label{canonical-details}
In this subsection we are going to derive the leading contribution of the complexity evolution in the canonical ensemble for arbitrary time.\\
First of all, let us notice that the integral \eqref{Phi(u)} can be evaluated analytically in the following two regimes
\begin{equation}
\Phi(u) \simeq 
\begin{cases*}
    \displaystyle 
    \frac{u}{\sqrt{2\pi}\,\beta^{3/2}} \, e^{\frac{2\pi^2}{\beta}}
    & $u \ll e^{\frac{2\pi^{2}}{\beta}}$, \\
    \displaystyle
    \frac{1}{(2\pi)^{3}}
    e^{2\pi\sqrt{\tfrac{2}{\beta}\log(u)}}
    \sqrt{\tfrac{2}{\beta}\log(u)}
    & $u \gg e^{\frac{2\pi^{2}}{\beta}}$.
\end{cases*}
\end{equation}
In what follows, we are going to prove that the kernel integral ${\cal I}'(0)$ \eqref{I0,can}, in the regime $u\leq u_{<}\ll e^{\frac{2\pi^{2}}{\beta}}$, turns out to give the main contribution to the complexity. \par
In the former small-$u$ range, we get straightforwardly
\begin{equation}\label{mainI'(0)}
    {\cal I}'(0)_{\text{main}}=\hat{y}(s)\,\frac{e^{-u_{<}(\hat{y}(s)+a_0\hat{Z}(\beta))}\left[a_0\hat{Z}(\beta)\left(-1+e^{u_{<}(\hat{y}(s)+a_0\hat{Z}(\beta))}\right)+u_{<}\hat{y}(s)\left(\hat{y}(s)+a_0\hat{Z}(\beta)\right)\right]}{\left(\hat{y}(s)+a_0\hat{Z}(\beta)\right)^{2}}\,,
\end{equation}
where
\begin{equation}
    \hat{Z}(\beta)=\frac{e^{\frac{2\pi^{2}}{\beta}}}{\sqrt{2\pi}\beta^{3/2}}\,,
\end{equation}
is the (un-normalized) disc-partition function evaluated in the saddle-point approximation. Plugging the kernel contribution \eqref{mainI'(0)} into eq.\,\eqref{early_time_comp_can} yields, upon performing a saddle-point approximation,
\begin{equation}\label{mainmain}
    \mathcal{C}(t)_{\text{main}}\simeq a_{0}t\sqrt{2\pi}\frac{e^{\frac{4\pi^{2}}{\beta}}}{\beta^{5/2}}\frac{a_0\hat{Z}(\beta)e^{\frac{2\pi^{2}}{\beta}}+e^{-a_0\hat{Z}(\beta)e^{\frac{2\pi^{2}}{\beta}}\beta}(-a_0\hat{Z}(\beta)+\beta e^{-\frac{2\pi^{2}}{\beta}})}{\left(1+a_0\hat{Z}(\beta)e^{\frac{2\pi^{2}}{\beta}}\right)^{2}}\,,
\end{equation}
where we chose $u_{<}=e^{\frac{2\pi^{2}}{\beta}}\beta\ll e^{\frac{2\pi^{2}}{\beta}}$.\\
Regarding the $u$-tail, let us notice that
\begin{equation}
    {\cal I}'(0)_{\text{tail}}=\int_{u_{>}}^{+\infty}du\,\left(1-u \hat{y}(s)\right) e^{-u \hat{y}(s)}e^{-a_0\Phi_{\text{tail}}(u)}\,,
\end{equation}
is dominated around the lower endpoint $u= u_{>}\gg e^{\frac{2\pi^{2}}{\beta}}$. Therefore, using the Laplace endpoint approximation, the above integral reduces to
\begin{equation}
    {\cal I}'(0)_{\text{tail}}\simeq \frac{\hat{y}(s)(1-u_{>}\hat{y}(s))}{\left(a_0\Phi'_{\text{tail}}(u_{>})+\hat{y}(s)\right)}e^{-u_{>}\hat{y}(s)-a_0\Phi_{\text{tail}}(u_{>})}\,,
\end{equation}
which, upon taking $u_{>}= \beta^{-1}e^{\frac{2\pi^{2}}{\beta}}\gg e^{\frac{2\pi^{2}}{\beta}}$ and performing the saddle point approximation, yields the following complexity tail contribution
\begin{equation}
    \mathcal{C}(t)_{\text{tail}}\simeq\, -a_{0}t\sqrt{2\pi}\frac{e^{\frac{4\pi^{2}}{\beta}}}{\beta^{5/2}}\frac{e^{-\frac{1}{\beta}-(2\pi)^{3/2}\beta^{1/2}a_0\hat{Z}(\beta)e^{\frac{2\pi^{2}}{\beta}}}}{\beta\left(1+(2\pi\beta)^{3/2}a_0\hat{Z}(\beta)e^{\frac{2\pi^{2}}{\beta}}\right)}\,.
\end{equation}
The complexity contribution due to the gap region $u_{<}<u<u_{>}$ can be computed by using the following uniform approximation for $\Phi(u)$,
\begin{equation}
    \Phi(u)=\int_{0}^{s_{*}}ds\,\rho(s) + u\int_{s_{*}}^{+\infty}ds\,\rho(s)\hat{y}(s)\,,
\end{equation}
where, in the gap region,
\begin{equation}
    s_{*}=\sqrt{\frac{2}{\beta}\log{u}}\simeq \frac{2\pi}{\beta}\,+\mathcal{O}(\beta\log{\beta})\,.
\end{equation}
Finally, one finds at leading order,
\begin{equation}
    \mathcal{C}(t)_{\text{gap}}\simeq a_{0}t\sqrt{2\pi}\frac{e^{\frac{4\pi^{2}}{\beta}}}{\beta^{5/2}}\frac{(a_0\hat{Z}(\beta)+e^{-\frac{2\pi^{2}}{\beta}})e^{-\frac{1}{\beta}(a_0\hat{Z}(\beta)e^{\frac{2\pi^{2}}{\beta}}+1)}}{\beta\left(1+a_0\hat{Z}(\beta)e^{\frac{2\pi^{2}}{\beta}}\right)^{2}}\,.
\end{equation}
It is straightforward to check that the leading-order contribution at all times is due to $\mathcal{C}(t)_{\text{main}}$ \eqref{mainmain}, explicitly  
\begin{equation}\label{dominant-canonical}
\boxed{
    \mathcal{C}(t)\simeq t\sqrt{2\pi}\frac{e^{\frac{4\pi^{2}}{\beta}}}{\beta^{5/2}}\frac{a_0^{2}\hat{Z}(\beta)e^{\frac{2\pi^{2}}{\beta}}}{\left(1+a_0\hat{Z}(\beta)e^{\frac{2\pi^{2}}{\beta}}\right)^{2}}=\frac{2\pi}{\beta}\,t\,\,\frac{e^{2\left(S_{0}+\frac{4\pi^{2}}{\beta}\right)}}{\left(\sqrt{2\pi}\,\beta^{3/2}\,k\,+\,e^{S_{0}+\frac{4\pi^{2}}{\beta}}\right)^{2}}\,}\,.
\end{equation}
One can argue that, at zeroth order in the $a_0\to0$ limit, eq.\,\eqref{dominant-canonical} does not capture a residual non-vanishing complexity displayed by both the tail and the gap contributions. However, this limit can be studied exactly through 
\begin{equation}
    {\cal I}'(0)=\hat{y}(s)\int_{0}^{+\infty}du\,\left(1-u \hat{y}(s)\right) e^{-u \hat{y}(s)}=0\,,
\end{equation}
leading us to conclude that eq.\,\eqref{dominant-canonical} captures all the complexity behaviour in the canonical ensemble.

\end{appendix}

\end{document}